\documentclass[journal,10pt,twoside,twocolumn]{IEEEtran}

\IEEEoverridecommandlockouts

\usepackage{algorithm}
\usepackage{algpseudocode}
\usepackage{amsmath,amsfonts} %citesort removed
\usepackage{amssymb, amsthm}
\usepackage{graphicx}
\usepackage{subfigure}
\usepackage{booktabs}
\usepackage{multirow}
\usepackage{mathrsfs}
\usepackage{cite}
\usepackage{units}
\usepackage{cancel}
\usepackage{upgreek}
\usepackage{color}
\DeclareGraphicsExtensions{.pdf,.jpeg,.png,.epbibdesks,.jpg}
\usepackage{epstopdf}
\usepackage{microtype}
\usepackage{balance}
\usepackage{paralist}
\usepackage{hyperref}
\usepackage{afterpage}
\usepackage{framed}
\usepackage{nicefrac}
\usepackage{adjustbox}
\usepackage{xspace}
\usepackage{bm}
\usepackage{fancyref}
\usepackage{textcomp}
\usepackage{stmaryrd}

\newcommand{\safemath}[2]{\newcommand{#1}{\ensuremath{#2}\xspace}}

\safemath{\bma}{\mathbf{a}}
\safemath{\bmb}{\mathbf{b}}
\safemath{\bmc}{\mathbf{c}}
\safemath{\bmd}{\mathbf{d}}
\safemath{\bme}{\mathbf{e}}
\safemath{\bmf}{\mathbf{f}}
\safemath{\bmg}{\mathbf{g}}
\safemath{\bmh}{\mathbf{h}}
\safemath{\bmi}{\mathbf{i}}
\safemath{\bmj}{\mathbf{j}}
\safemath{\bmk}{\mathbf{k}}
\safemath{\bml}{\mathbf{l}}
\safemath{\bmm}{\mathbf{m}}
\safemath{\bmn}{\mathbf{n}}
\safemath{\bmo}{\mathbf{o}}
\safemath{\bmp}{\mathbf{p}}
\safemath{\bmq}{\mathbf{q}}
\safemath{\bmr}{\mathbf{r}}
\safemath{\bms}{\mathbf{s}}
\safemath{\bmt}{\mathbf{t}}
\safemath{\bmu}{\mathbf{u}}
\safemath{\bmv}{\mathbf{v}}
\safemath{\bmw}{\mathbf{w}}
\safemath{\bmx}{\mathbf{x}}
\safemath{\bmy}{\mathbf{y}}
\safemath{\bmz}{\mathbf{z}}
\safemath{\bmzero}{\mathbf{0}}
\safemath{\bmone}{\mathbf{1}}

\bmdefine{\biad}{a}
\bmdefine{\bibd}{b}
\bmdefine{\bicd}{c}
\bmdefine{\bidd}{d}
\bmdefine{\bied}{e}
\bmdefine{\bifd}{f}
\bmdefine{\bigd}{g}
\bmdefine{\bihd}{h}
\bmdefine{\biid}{i}
\bmdefine{\bijd}{j}
\bmdefine{\bikd}{k}
\bmdefine{\bild}{l}
\bmdefine{\bimd}{m}
\bmdefine{\bind}{n}
\bmdefine{\biod}{o}
\bmdefine{\bipd}{p}
\bmdefine{\biqd}{q}
\bmdefine{\bird}{r}
\bmdefine{\bisd}{s}
\bmdefine{\bitd}{t}
\bmdefine{\biud}{u}
\bmdefine{\bivd}{v}
\bmdefine{\biwd}{w}
\bmdefine{\bixd}{x}
\bmdefine{\biyd}{y}
\bmdefine{\bizd}{z}

\bmdefine{\bixid}{\xi}
\bmdefine{\bilambdad}{\lambda}
\bmdefine{\bimud}{\mu}
\bmdefine{\bithetad}{\theta}
\bmdefine{\biphid}{\phi}
\bmdefine{\bideltad}{\delta}

\safemath{\bmia}{\biad}
\safemath{\bmib}{\bibd}
\safemath{\bmic}{\bicd}
\safemath{\bmid}{\bidd}
\safemath{\bmie}{\bied}
\safemath{\bmif}{\bifd}
\safemath{\bmig}{\bigd}
\safemath{\bmih}{\bihd}
\safemath{\bmii}{\biid}
\safemath{\bmij}{\bijd}
\safemath{\bmik}{\bikd}
\safemath{\bmil}{\bild}
\safemath{\bmim}{\bimd}
\safemath{\bmin}{\bind}
\safemath{\bmio}{\biod}
\safemath{\bmip}{\bipd}
\safemath{\bmiq}{\biqd}
\safemath{\bmir}{\bird}
\safemath{\bmis}{\bisd}
\safemath{\bmit}{\bitd}
\safemath{\bmiu}{\biud}
\safemath{\bmiv}{\bivd}
\safemath{\bmiw}{\biwd}
\safemath{\bmix}{\bixd}
\safemath{\bmiy}{\biyd}
\safemath{\bmiz}{\bizd}

\safemath{\bmxi}{\bixid}
\safemath{\bmlambda}{\bilambdad}
\safemath{\bmmu}{\bimud}
\safemath{\bmtheta}{\bithetad}
\safemath{\bmphi}{\biphid}
\safemath{\bmdelta}{\bideltad}

\safemath{\bA}{\mathbf{A}}
\safemath{\bB}{\mathbf{B}}
\safemath{\bC}{\mathbf{C}}
\safemath{\bD}{\mathbf{D}}
\safemath{\bE}{\mathbf{E}}
\safemath{\bF}{\mathbf{F}}
\safemath{\bG}{\mathbf{G}}
\safemath{\bH}{\mathbf{H}}
\safemath{\bI}{\mathbf{I}}
\safemath{\bJ}{\mathbf{J}}
\safemath{\bK}{\mathbf{K}}
\safemath{\bL}{\mathbf{L}}
\safemath{\bM}{\mathbf{M}}
\safemath{\bN}{\mathbf{N}}
\safemath{\bO}{\mathbf{O}}
\safemath{\bP}{\mathbf{P}}
\safemath{\bQ}{\mathbf{Q}}
\safemath{\bR}{\mathbf{R}}
\safemath{\bS}{\mathbf{S}}
\safemath{\bT}{\mathbf{T}}
\safemath{\bU}{\mathbf{U}}
\safemath{\bV}{\mathbf{V}}
\safemath{\bW}{\mathbf{W}}
\safemath{\bX}{\mathbf{X}}
\safemath{\bY}{\mathbf{Y}}
\safemath{\bZ}{\mathbf{Z}}

\safemath{\bZero}{\mathbf{0}}
\safemath{\bOne}{\mathbf{1}}
\safemath{\bDelta}{\mathbf{\Delta}}
\safemath{\bLambda}{\mathbf{\UpLambda}}
\safemath{\bPhi}{\mathbf{\Upphi}}
\safemath{\bSigma}{\mathbf{\Upsigma}}
\safemath{\bOmega}{\mathbf{\Upomega}}
\safemath{\bTheta}{\mathbf{\Uptheta}}

\bmdefine{\biAd}{A}
\bmdefine{\biBd}{B}
\bmdefine{\biCd}{C}
\bmdefine{\biDd}{D}
\bmdefine{\biEd}{E}
\bmdefine{\biFd}{F}
\bmdefine{\biGd}{G}
\bmdefine{\biHd}{H}
\bmdefine{\biId}{I}
\bmdefine{\biJd}{J}
\bmdefine{\biKd}{K}
\bmdefine{\biLd}{L}
\bmdefine{\biMd}{M}
\bmdefine{\biOd}{N}
\bmdefine{\biPd}{O}
\bmdefine{\biQd}{P}
\bmdefine{\biRd}{R}
\bmdefine{\biSd}{S}
\bmdefine{\biTd}{T}
\bmdefine{\biUd}{U}
\bmdefine{\biVd}{V}
\bmdefine{\biWd}{W}
\bmdefine{\biXd}{X}
\bmdefine{\biYd}{Y}
\bmdefine{\biZd}{Z}

\bmdefine{\biDelta}{\Delta}
\bmdefine{\biLambda}{\Lambda}
\bmdefine{\biPhi}{\Phi}
\bmdefine{\biSigma}{\Sigma}
\bmdefine{\biOmega}{\Omega}
\bmdefine{\biTheta}{\Theta}

\safemath{\bimA}{\biAd}
\safemath{\bimB}{\biBd}
\safemath{\bimC}{\biCd}
\safemath{\bimD}{\biDd}
\safemath{\bimE}{\biEd}
\safemath{\bimF}{\biFd}
\safemath{\bimG}{\biGd}
\safemath{\bimH}{\biHd}
\safemath{\bimI}{\biId}
\safemath{\bimJ}{\biJd}
\safemath{\bimK}{\biKd}
\safemath{\bimL}{\biLd}
\safemath{\bimM}{\biMd}
\safemath{\bimN}{\biNd}
\safemath{\bimO}{\biOd}
\safemath{\bimP}{\biPd}
\safemath{\bimQ}{\biQd}
\safemath{\bimR}{\biRd}
\safemath{\bimS}{\biSd}
\safemath{\bimT}{\biTd}
\safemath{\bimU}{\biUd}
\safemath{\bimV}{\biVd}
\safemath{\bimW}{\biWd}
\safemath{\bimX}{\biXd}
\safemath{\bimY}{\biYd}
\safemath{\bimZ}{\biZd}

\safemath{\bimDelta}{\biDelta}
\safemath{\bimLambda}{\biLambda}
\safemath{\bimPhi}{\biPhi}
\safemath{\bimSigma}{\biSigma}
\safemath{\bimOmega}{\biOmega}
\safemath{\bimTheta}{\biTheta}

\safemath{\setA}{\mathcal{A}}
\safemath{\setB}{\mathcal{B}}
\safemath{\setC}{\mathcal{C}}
\safemath{\setD}{\mathcal{D}}
\safemath{\setE}{\mathcal{E}}
\safemath{\setF}{\mathcal{F}}
\safemath{\setG}{\mathcal{G}}
\safemath{\setH}{\mathcal{H}}
\safemath{\setI}{\mathcal{I}}
\safemath{\setJ}{\mathcal{J}}
\safemath{\setK}{\mathcal{K}}
\safemath{\setL}{\mathcal{L}}
\safemath{\setM}{\mathcal{M}}
\safemath{\setN}{\mathcal{N}}
\safemath{\setO}{\mathcal{O}}
\safemath{\setP}{\mathcal{P}}
\safemath{\setQ}{\mathcal{Q}}
\safemath{\setR}{\mathcal{R}}
\safemath{\setS}{\mathcal{S}}
\safemath{\setT}{\mathcal{T}}
\safemath{\setU}{\mathcal{U}}
\safemath{\setV}{\mathcal{V}}
\safemath{\setW}{\mathcal{W}}
\safemath{\setX}{\mathcal{X}}
\safemath{\setY}{\mathcal{Y}}
\safemath{\setZ}{\mathcal{Z}}
\safemath{\emptySet}{\varnothing}

\safemath{\colA}{\mathscr{A}}
\safemath{\colB}{\mathscr{B}}
\safemath{\colC}{\mathscr{C}}
\safemath{\colD}{\mathscr{D}}
\safemath{\colE}{\mathscr{E}}
\safemath{\colF}{\mathscr{F}}
\safemath{\colG}{\mathscr{G}}
\safemath{\colH}{\mathscr{H}}
\safemath{\colI}{\mathscr{I}}
\safemath{\colJ}{\mathscr{J}}
\safemath{\colK}{\mathscr{K}}
\safemath{\colL}{\mathscr{L}}
\safemath{\colM}{\mathscr{M}}
\safemath{\colN}{\mathscr{N}}
\safemath{\colO}{\mathscr{O}}
\safemath{\colP}{\mathscr{P}}
\safemath{\colQ}{\mathscr{Q}}
\safemath{\colR}{\mathscr{R}}
\safemath{\colS}{\mathscr{S}}
\safemath{\colT}{\mathscr{T}}
\safemath{\colU}{\mathscr{U}}
\safemath{\colV}{\mathscr{V}}
\safemath{\colW}{\mathscr{W}}
\safemath{\colX}{\mathscr{X}}
\safemath{\colY}{\mathscr{Y}}
\safemath{\colZ}{\mathscr{Z}}

\safemath{\opA}{\mathbb{A}}
\safemath{\opB}{\mathbb{B}}
\safemath{\opC}{\mathbb{C}}
\safemath{\opD}{\mathbb{D}}
\safemath{\opE}{\mathbb{E}}
\safemath{\opF}{\mathbb{F}}
\safemath{\opG}{\mathbb{G}}
\safemath{\opH}{\mathbb{H}}
\safemath{\opI}{\mathbb{I}}
\safemath{\opJ}{\mathbb{J}}
\safemath{\opK}{\mathbb{K}}
\safemath{\opL}{\mathbb{L}}
\safemath{\opM}{\mathbb{M}}
\safemath{\opN}{\mathbb{N}}
\safemath{\opO}{\mathbb{O}}
\safemath{\opP}{\mathbb{P}}
\safemath{\opQ}{\mathbb{Q}}
\safemath{\opR}{\mathbb{R}}
\safemath{\opS}{\mathbb{S}}
\safemath{\opT}{\mathbb{T}}
\safemath{\opU}{\mathbb{U}}
\safemath{\opV}{\mathbb{V}}
\safemath{\opW}{\mathbb{W}}
\safemath{\opX}{\mathbb{X}}
\safemath{\opY}{\mathbb{Y}}
\safemath{\opZ}{\mathbb{Z}}
\safemath{\opZero}{\mathbb{O}}
\safemath{\identityop}{\opI}

\safemath{\veca}{\bma}
\safemath{\vecb}{\bmb}
\safemath{\vecc}{\bmc}
\safemath{\vecd}{\bmd}
\safemath{\vece}{\bme}
\safemath{\vecf}{\bmf}
\safemath{\vecg}{\bmg}
\safemath{\vech}{\bmh}
\safemath{\veci}{\bmi}
\safemath{\vecj}{\bmj}
\safemath{\veck}{\bmk}
\safemath{\vecl}{\bml}
\safemath{\vecm}{\bmm}
\safemath{\vecn}{\bmn}
\safemath{\veco}{\bmo}
\safemath{\vecp}{\bmp}
\safemath{\vecq}{\bmq}
\safemath{\vecr}{\bmr}
\safemath{\vecs}{\bms}
\safemath{\vect}{\bmt}
\safemath{\vecu}{\bmu}
\safemath{\vecv}{\bmv}
\safemath{\vecw}{\bmw}
\safemath{\vecx}{\bmx}
\safemath{\vecy}{\bmy}
\safemath{\vecz}{\bmz}

\safemath{\veczero}{\bmzero}
\safemath{\vecone}{\bmone}
\safemath{\vecxi}{\bmxi}
\safemath{\veclambda}{\bmlambda}
\safemath{\vecmu}{\bmmu}
\safemath{\vectheta}{\bmtheta}
\safemath{\vecphi}{\bmphi}
\safemath{\vecdelta}{\bmdelta}

\safemath{\matA}{\bA}
\safemath{\matB}{\bB}
\safemath{\matC}{\bC}
\safemath{\matD}{\bD}
\safemath{\matE}{\bE}
\safemath{\matF}{\bF}
\safemath{\matG}{\bG}
\safemath{\matH}{\bH}
\safemath{\matI}{\bI}
\safemath{\matJ}{\bJ}
\safemath{\matK}{\bK}
\safemath{\matL}{\bL}
\safemath{\matM}{\bM}
\safemath{\matN}{\bN}
\safemath{\matO}{\bO}
\safemath{\matP}{\bP}
\safemath{\matQ}{\bQ}
\safemath{\matR}{\bR}
\safemath{\matS}{\bS}
\safemath{\matT}{\bT}
\safemath{\matU}{\bU}
\safemath{\matV}{\bV}
\safemath{\matW}{\bW}
\safemath{\matX}{\bX}
\safemath{\matY}{\bY}
\safemath{\matZ}{\bZ}
\safemath{\matzero}{\bmzero}

\safemath{\matDelta}{\bDelta}
\safemath{\matLambda}{\bLambda}
\safemath{\matPhi}{\bPhi}
\safemath{\matSigma}{\bSigma}
\safemath{\matOmega}{\bOmega}
\safemath{\matTheta}{\bTheta}

\safemath{\matidentity}{\matI}
\safemath{\matone}{\matO}

\safemath{\rnda}{A}
\safemath{\rndb}{B}
\safemath{\rndc}{C}
\safemath{\rndd}{D}
\safemath{\rnde}{E}
\safemath{\rndf}{F}
\safemath{\rndg}{G}
\safemath{\rndh}{H}
\safemath{\rndi}{I}
\safemath{\rndj}{J}
\safemath{\rndk}{K}
\safemath{\rndl}{L}
\safemath{\rndm}{M}
\safemath{\rndn}{N}
\safemath{\rndo}{O}
\safemath{\rndp}{P}
\safemath{\rndq}{Q}
\safemath{\rndr}{R}
\safemath{\rnds}{S}
\safemath{\rndt}{T}
\safemath{\rndu}{U}
\safemath{\rndv}{V}
\safemath{\rndw}{W}
\safemath{\rndx}{X}
\safemath{\rndy}{Y}
\safemath{\rndz}{Z}

\safemath{\rveca}{\bimA}
\safemath{\rvecb}{\bimB}
\safemath{\rvecc}{\bimC}
\safemath{\rvecd}{\bimD}
\safemath{\rvece}{\bimE}
\safemath{\rvecf}{\bimF}
\safemath{\rvecg}{\bimG}
\safemath{\rvech}{\bimH}
\safemath{\rveci}{\bimI}
\safemath{\rvecj}{\bimJ}
\safemath{\rveck}{\bimK}
\safemath{\rvecl}{\bimL}
\safemath{\rvecm}{\bimM}
\safemath{\rvecn}{\bimN}
\safemath{\rveco}{\bomO}
\safemath{\rvecp}{\bimP}
\safemath{\rvecq}{\bimQ}
\safemath{\rvecr}{\bimR}
\safemath{\rvecs}{\bimS}
\safemath{\rvect}{\bimT}
\safemath{\rvecu}{\bimU}
\safemath{\rvecv}{\bimV}
\safemath{\rvecw}{\bimW}
\safemath{\rvecx}{\bimX}
\safemath{\rvecy}{\bimY}
\safemath{\rvecz}{\bimZ}

\safemath{\rvecxi}{\bmxi}
\safemath{\rveclambda}{\bmlambda}
\safemath{\rvecmu}{\bmmu}
\safemath{\rvectheta}{\bmtheta}
\safemath{\rvecphi}{\bmphi}

\safemath{\rmatA}{\bimA}
\safemath{\rmatB}{\bimB}
\safemath{\rmatC}{\bimC}
\safemath{\rmatD}{\bimD}
\safemath{\rmatE}{\bimE}
\safemath{\rmatF}{\bimF}
\safemath{\rmatG}{\bimG}
\safemath{\rmatH}{\bimH}
\safemath{\rmatI}{\bimI}
\safemath{\rmatJ}{\bimJ}
\safemath{\rmatK}{\bimK}
\safemath{\rmatL}{\bimL}
\safemath{\rmatM}{\bimM}
\safemath{\rmatN}{\bimN}
\safemath{\rmatO}{\bimO}
\safemath{\rmatP}{\bimP}
\safemath{\rmatQ}{\bimQ}
\safemath{\rmatR}{\bimR}
\safemath{\rmatS}{\bimS}
\safemath{\rmatT}{\bimT}
\safemath{\rmatU}{\bimU}
\safemath{\rmatV}{\bimV}
\safemath{\rmatW}{\bimW}
\safemath{\rmatX}{\bimX}
\safemath{\rmatY}{\bimY}
\safemath{\rmatZ}{\bimZ}

\safemath{\rmatDelta}{\bimDelta}
\safemath{\rmatLambda}{\bimLambda}
\safemath{\rmatPhi}{\bimPhi}
\safemath{\rmatSigma}{\bimSigma}
\safemath{\rmatOmega}{\bimOmega}
\safemath{\rmatTheta}{\bimTheta}

\newenvironment{textbmatrix}{	\setlength{\arraycolsep}{2.5pt}%
								\big[\begin{matrix}}{\end{matrix}\big]%
								\raisebox{0.08ex}{\vphantom{M}}}
								
\def\be{\begin{equation}}
\def\ee{\end{equation}}
\def\een{\nonumber \end{equation}}
\def\mat{\begin{bmatrix}}
\def\emat{\end{bmatrix}}
\def\btm{\begin{textbmatrix}}
\def\etm{\end{textbmatrix}}

\def\ba#1\ea{\begin{align}#1\end{align}}
\def\bas#1\eas{\begin{align*}#1\end{align*}}
\def\bs#1\es{\begin{split}#1\end{split}} 
\def\bg#1\eg{\begin{gather}#1\end{gather}}
\def\bml#1\eml{\begin{multline}#1\end{multline}}
\def\bi#1\ei{\begin{itemize}#1\end{itemize}} 

\DeclareMathOperator*{\argmin}{arg\;min}		% arg min
\DeclareMathOperator*{\argmax}{arg\;max}		% arg max
\newcommand{\opt}[1]{\ensuremath{#1^{*}}} 	% optimum parameter
\safemath{\dirac}{\delta}					% Dirac delta
\safemath{\krond}{\dirac}					% Kronecker delta
\safemath{\upto}{\uparrow}
\safemath{\downto}{\downarrow}
\safemath{\iu}{j}							% imaginary unit
\safemath{\ev}{\lambda}						% eigenvalue
\safemath{\hilseqspace}{l^{2}}				% Hilbert sequence space
\newcommand{\banachfunspace}[1]{\setL^{#1}}	% Banach function space
\safemath{\hilfunspace}{\banachfunspace{2}}	% Hilbert function space
\safemath{\SNR}{\textsf{SNR}} 				% signal to noise ratio
\safemath{\PAR}{\textsf{PAR}} 				% signal to noise ratio
\safemath{\No}{N_0}							% noise spectral density
\safemath{\Es}{E_s}							% energy per symbol
\safemath{\Eb}{E_b}							% energy per bit
\safemath{\EbNo}{\frac{\Eb}{\No}}
\safemath{\EsNo}{\frac{\Es}{\No}}

\DeclareMathOperator{\CHop}{\ensuremath{\opH}} % channel operator
\safemath{\tvir}{\rndh_{\CHop}}				% time-varying impulse response
\safemath{\tvtf}{\rndl_{\CHop}}				% 	-''- transfer function
\safemath{\spf}{\rnds_{\CHop}}				% spreading function
\safemath{\bff}{H_{\CHop}}					% bi-freuqency function

\safemath{\ircf}{r_{h}}						% impulse response correlation fn.
\safemath{\tftvcf}{r_{s}}					% scattering function
\safemath{\tfcf}{r_{l}}						% time-frequency correlation fn.
\safemath{\bfcf}{r_{H}}						% bi-frequency correlation fn.

\safemath{\tcorr}{c_h}						% time-correlation function
\safemath{\scf}{c_{s}}						% spreading function
\safemath{\tfcorr}{c_{l}}					% transfer-function correlation
\safemath{\fcorr}{c_{H}}						% frequency-correlation function

\safemath{\mi}{I}							% mutual information
\safemath{\capacity}{C}						% capacity

\safemath{\normal}{\mathcal{N}}			% normal distribution
\safemath{\jpg}{\mathcal{CN}}			% jointly proper Gaussian
\safemath{\mchain}{\leftrightarrow}		% Markov chain
\safemath{\dB}{\,\mathrm{dB}}
\safemath{\dBm}{\,\mathrm{dBm}}
\safemath{\Hz}{\,\mathrm{Hz}}
\safemath{\kHz}{\,\mathrm{kHz}}
\safemath{\MHz}{\,\mathrm{MHz}}
\safemath{\GHz}{\,\mathrm{GHz}}
\safemath{\s}{\,\mathrm{s}}
\safemath{\ms}{\,\mathrm{ms}}
\safemath{\mus}{\,\mathrm{\text{\textmu}s}}
\safemath{\ns}{\,\mathrm{ns}}
\safemath{\ps}{\,\mathrm{ps}}
\safemath{\meter}{\,\mathrm{m}}
\safemath{\mm}{\,\mathrm{mm}}
\safemath{\cm}{\,\mathrm{cm}}
\safemath{\m}{\,\mathrm{m}}
\safemath{\W}{\,\mathrm{W}}
\safemath{\mW}{\, \mathrm{mW}}
\safemath{\J}{\,\mathrm{J}}
\safemath{\K}{\,\mathrm{K}}
\safemath{\bit}{\,\mathrm{bit}}
\safemath{\nat}{\,\mathrm{nat}}

\safemath{\define}{\triangleq}			% definition

\safemath{\equivalent}{\sim}
\safemath{\distas}{\sim}					% distributed according to
\safemath{\sdiff}{\Delta}				% symmetric set difference

\safemath{\reals}{\mathbb{R}}
\safemath{\positivereals}{\reals_{+}}
\safemath{\integers}{\mathbb{Z}}
\safemath{\posint}{\integers_{+}}
\safemath{\naturals}{\mathbb{N}}
\safemath{\posnaturals}{\naturals_{+}}
\safemath{\complexset}{\mathbb{C}}
\safemath{\rationals}{\mathbb{Q}}

\newcommand*{\fancyrefapplabelprefix}{app}		% Appendix
\newcommand*{\fancyrefthmlabelprefix}{thm}		% Theorem
\newcommand*{\fancyreflemlabelprefix}{lem}		% Lemma
\newcommand*{\fancyrefcorlabelprefix}{cor}		% Corollary
\newcommand*{\fancyrefdeflabelprefix}{def}		% Definition
\newcommand*{\fancyrefproplabelprefix}{prop}	% Proposition
\newcommand*{\fancyrefobslabelprefix}{obs}		% Observation 
\newcommand*{\fancyrefalglabelprefix}{alg}		% Algorithm
\newcommand*{\fancyrefasmlabelprefix}{asm}	    % Assumption
\newcommand*{\fancyreftbllabelprefix}{tbl}	    % Assumption

\frefformat{vario}{\fancyrefseclabelprefix}{Section~#1}
\frefformat{vario}{\fancyrefthmlabelprefix}{Theorem~#1}
\frefformat{vario}{\fancyreflemlabelprefix}{Lemma~#1}
\frefformat{vario}{\fancyrefcorlabelprefix}{Corollary~#1}
\frefformat{vario}{\fancyrefdeflabelprefix}{Definition~#1}
\frefformat{vario}{\fancyrefobslabelprefix}{Observation~#1}
\frefformat{vario}{\fancyrefasmlabelprefix}{Assumption~#1}
\frefformat{vario}{\fancyreffiglabelprefix}{Figure~#1}
\frefformat{vario}{\fancyrefapplabelprefix}{Appendix~#1} 
\frefformat{vario}{\fancyrefproplabelprefix}{Proposition~#1}
\frefformat{vario}{\fancyrefalglabelprefix}{Algorithm~#1}
\frefformat{vario}{\fancyrefeqlabelprefix}{(#1)}
\frefformat{vario}{\fancyreftbllabelprefix}{Table~#1}

\newtheorem{thm}{Theorem}
\newtheorem{rem}{Remark}

\safemath{\dictab}{[\,\dicta\,\,\dictb\,]}

\safemath{\ysig}{\bmy}
\safemath{\ysighat}{\hat{\ysig}}
\safemath{\ysigdim}{M}
\safemath{\xsig}{\bmx}
\safemath{\xsigdim}{N}
\safemath{\nx}{n_x}
\safemath{\zsig}{\bmz}
\safemath{\zsigdim}{\ysigdim}
\safemath{\rsig}{\bmr}
\safemath{\Adict}{\bA}
\safemath{\Adicttilde}{\widetilde{\Adict}}
\safemath{\Adictdim}{\outputdim\times\xsigdim}
\safemath{\avec}{\bma}
\safemath{\avectilde}{\tilde{\avec}}
\safemath{\Bdict}{\bB}
\safemath{\Bdicttilde}{\widetilde{\Bdict}}
\safemath{\Cdict}{\bC}
\safemath{\cvec}{\bmc}
\safemath{\Ddict}{\bD}
\safemath{\Ddictdim}{\ysigdim\times\xsigdim}
\safemath{\dvec}{\bmd}
\safemath{\Ddicttilde}{\widetilde{\bD}}
\safemath{\Bonb}{\bB}
\safemath{\bvec}{\bmb}
\safemath{\Bonbdim}{\ysigdim\times\ysigdim}
\safemath{\noise}{\bmn}
\safemath{\noisedim}{\ysigim}
\safemath{\err}{\bme}
\safemath{\errdim}{\ysigdim}
\safemath{\errset}{\setE}
\safemath{\nerr}{n_e}
\safemath{\delop}{\bP_\errset}
\safemath{\delopc}{\bP_{{\errset}^c}}

\safemath{\cplxi}{\imath}
\safemath{\cplxj}{\jmath}
\safemath{\dict}{\matD}
\safemath{\inputdim}{N}		% number of columns of dictionary D
\safemath{\outputdim}{M}		%number of rows of dictionary D
\safemath{\sparsity}{S}	%sparsity
\safemath{\inputdimA}{{N_a}}	%total number of elements in dictionary A
\safemath{\inputdimB}{{N_b}}	%total number of elements in dictionary B
\safemath{\elemA}{{n_a}}	%number of elements chosen from dictionary A
\safemath{\elemB}{{n_b}}	%number of elements chosen from dictionary B
\safemath{\resA}{\matR_a}	%restriction map to elements of dictionary A
\safemath{\resB}{\matR_b}	%restriction map to elements of dictionary B
\safemath{\subD}{\matS} %subdictionary
\safemath{\subA}{\matS_a} %subdictionary part of A
\safemath{\subB}{\matS_b} %subdictionary part of B
\safemath{\dicta}{\matA} 	% first subdictionary
\safemath{\dictb}{\matB} 	% second subdictionary
\safemath{\hollowS}{H}
\safemath{\hollowA}{H_a}
\safemath{\hollowB}{H_b}
\safemath{\cross}{Z}
\safemath{\coh}{\mu_d}			% coherence dictionary
\safemath{\coha}{\mu_a}			% coherence first subdictionary
\safemath{\cohb}{\mu_b}			% coherence second subdictionary
\safemath{\mubs}{\nu}	%block sub-coherence
\safemath{\cohm}{\mu_m} %mutual coherence
\safemath{\dictset}{\setD}	% set of dictionaries
\safemath{\dictsetp}{\dictset(\coh,\coha,\cohb)}	% set of dictionaries parametrized
\safemath{\dictsetgen}{\dictset_\text{gen}}
\safemath{\dictsetgenp}{\dictsetgen(\coh)}
\safemath{\dictsetonb}{\dictset_\text{onb}}
\safemath{\dictsetonbp}{\dictsetonb(\coh)}

\safemath{\leftside}{U}
\safemath{\rightsideA}{R_a}
\safemath{\rightsideB}{R_b}

\safemath{\indexS}{\setI_S} %set of indices participating in sub-dictionary S

\safemath{\na}{n_a}			% cardinality of set of linearly independent columns of first dictionary
\safemath{\nb}{n_b}			% cardinality of set of linearly independent columns of second dictionary
\safemath{\coeffa}{p_i}	%coefficients for columns of A
\safemath{\coeffb}{q_j}	%coefficients for columns of B
\safemath{\seta}{\setP}		% set of linearly independent columns of A
\safemath{\setb}{\setQ}     % set of linearly independent columns of B
\safemath{\setw}{\setW}	%set of n largest elements of w
\safemath{\setz}{\setZ}	%set of L-n largest elements of z
\safemath{\cola}{\veca}		% generic element of the dictionary A
\safemath{\colb}{\vecb}		% generic element of the dictionary B
\safemath{\cold}{\vecd}		% generic element of the dictionary D
\safemath{\inputvec}{\vecx} 	%coefficient vector (input)
\safemath{\error}{\vece}	%error vector
\safemath{\noiseout}{\vecz} 	%noisy output vector
\safemath{\inputvecel}{x}
\safemath{\inputveca}{\vecx_a}
\safemath{\inputvecb}{\vecx_b}
\safemath{\outputvec}{\vecy}	%output of Dictionary
\safemath{\lambdamin}{\lambda_{\mathrm{min}}}
\safemath{\elltwo}{\ell_2}
\safemath{\ellone}{\ell_1}
\safemath{\ellzero}{\ell_0}
\safemath{\ellinf}{\ell_\infty}
\safemath{\ellinftilde}{\ell_{\widetilde\infty}}
\safemath{\licard}{Z(\coh,\coha,\cohb)}
\safemath{\xsol}{\hat{x}}
\safemath{\xbord}{x_b}		%Solution at the border
\safemath{\xstat}{x_s}		%Solution stationary in l0 prob
\safemath{\xstatLone}{\tilde{x}_s}
\safemath{\order}{\mathcal{O}} %order notation (big O)
\safemath{\scales}{\Theta} %scales as
\safemath{\ones}{\mathbf{1}} %all ones matrix
\safemath{\zeroes}{\mathbf{0}} %all zeroes matrix
\safemath{\thlone}{\kappa(\coh,\cohb)} %treshold l1 problem
\safemath{\constoneA}{\delta} %constant in l1 theorem to save space
\safemath{\constoneB}{\epsilon} %constant in l1 theorem to save space
\safemath{\nlarge}{L}				   %num large elements
\safemath{\sumlarge}{S_\nlarge}
\safemath{\maxlarger}{P_\nlarge}	   % maximum in Gribonval and Nielsen
\safemath{\Pzero}{\textrm{P0}}	
\safemath{\Pone}{\textrm{P1}}
\safemath{\vecfir}{\vecw}			 % \vecv element of the kernel of the dictionary, \vecv=[\vecfir \vecsec]
\safemath{\vecsec}{\vecz}
\safemath{\elvecfir}{w}              % element of vecfir
\safemath{\elvecsec}{z}				 % element of vecsec
\safemath{\nlargefir}{n}
\safemath{\normout}{\gamma}
\safemath{\auxfun}{h}
\safemath{\supp}{\textrm{supp}}%support

\safemath{\indexa}{\ell}
\safemath{\indexb}{r}
\safemath{\indexc}{i}
\safemath{\indexd}{j}

\safemath{\project}{P}%projector

\newcommand{\MR}{{B}}
\newcommand{\C}{\mathbb{C}}
\newcommand{\R}{\mathbb{R}}

\renewcommand{\bml}{\ensuremath{\boldsymbol \ell}}

\newcommand{\rev}[1]{\textcolor{black}{#1}}
\newcommand{\algname}{PrOX}
\newcommand{\apalgname}{APrOX}

\newtheorem{alg}{Algorithm}

\begin{document}
\title{VLSI Designs for Joint Channel Estimation and Data Detection in Large SIMO Wireless Systems} 

\author{
\IEEEauthorblockN{Oscar Casta\~neda, Tom Goldstein, and Christoph Studer} 
\thanks{O. Casta\~neda and C. Studer are with the School~of Electrical and Computer Engineering, Cornell University, Ithaca, NY; e-mail: oc66@cornell.edu, studer@cornell.edu; \rev{web: \url{http://vip.ece.cornell.edu}}} 
\thanks{T. Goldstein is with the Department of Computer Science, University of Maryland, College Park, MD; e-mail: tomg@cs.umd.edu}
\thanks{A short version of this paper summarizing the PrOX FPGA design has been presented at the IEEE \rev{Int.}~Symp.~on Circuits and Systems (ISCAS) 2017 \cite{CGC17}.}
\thanks{\rev{The MATLAB simulator for PrOX used in this paper is available on GitHub: \url{https://github.com/VIP-Group/PrOX}.}}
}

\maketitle

\begin{abstract}
Channel estimation errors have a critical impact on the reliability of wireless communication systems. 
While virtually all existing wireless receivers separate channel estimation from data detection, it is well known that joint channel estimation and data detection (JED) significantly outperforms conventional methods \rev{at the cost of high computational complexity.}
In this paper, we propose a novel JED algorithm and corresponding VLSI designs for large single-input multiple-output (SIMO) wireless systems that use constant-modulus constellations.
The proposed algorithm is referred to as PRojection Onto conveX hull (\algname) and relies on biconvex relaxation (BCR), which enables us to efficiently compute an approximate solution of the maximum-likelihood JED problem.
Since BCR solves a biconvex problem via alternating optimization, we provide a theoretical convergence analysis for PrOX.
We design a scalable, high-throughput VLSI architecture that uses a linear array of processing elements to minimize hardware complexity. We develop corresponding field-programmable gate array (FPGA) and application-specific integrated circuit (ASIC) designs, and we demonstrate that  \algname{} significantly outperforms the only other existing JED design in terms of throughput, hardware-efficiency, and energy-efficiency. 
\end{abstract}

% ABSTRACT FOR COPY PASTE:
% Channel estimation errors have a critical impact on the reliability of wireless communication systems. While virtually all existing wireless receivers separate channel estimation from data detection, it is well known that joint channel estimation and data detection (JED) significantly outperforms conventional methods at the cost of high computational complexity. In this paper, we propose a novel JED algorithm and corresponding VLSI designs for large single-input multiple-output (SIMO) wireless systems that use constant-modulus constellations. The proposed algorithm is referred to as PRojection Onto conveX hull (PrOX) and relies on biconvex relaxation (BCR), which enables us to efficiently compute an approximate solution of the maximum-likelihood JED problem. Since BCR solves a biconvex problem via alternating optimization, we provide a theoretical convergence analysis for PrOX. We design a scalable, high-throughput VLSI architecture that uses a linear array of processing elements to minimize hardware complexity. We develop corresponding field-programmable gate array (FPGA) and application-specific integrated circuit (ASIC) designs, and we demonstrate that  PrOX significantly outperforms the only other existing JED design in terms of throughput, hardware-efficiency, and energy-efficiency. 

%

%%%%%%%%%%%%%%%%%%%%%%%%%%%%%%%%%%%%%%%%%%%%%%%%%%%%

\section{Introduction}
\label{sec:intro}
\IEEEPARstart{W}{ireless} communication with a large number of antennas at the base-station (BS) will play a major role in fifth-generation (5G) systems. 
By equipping the BS with hundreds or thousands of antenna elements, large wireless systems enable fine-grained beamforming and, hence, improved spectral-efficiency within each cell compared to traditional communication systems that use a small number of BS antennas~\cite{Marzetta2010,Rusek2012,hoydis2013massive,larsson2014massive,andrews2014will,LUetal2014}.
All these advantages come at significantly increased signal-processing complexity at the BS~\cite{Wu2014}, which necessitates the development of high-performance transceiver algorithms that scale well to large BS array sizes and can be implemented  in very-large scale integration (VLSI) circuits at low costs and in an energy-efficient manner.

\subsection{The Importance of Accurate Channel State Information}
Due to the fine-grained nature of beamforming in such large wireless systems, the acquisition of accurate channel state information (CSI) at the BS is critical for reliable, \rev{high-throughput} data transmission.
In particular, accurate channel state information is not only required in the uplink (to coherently detect data transmitted from the users to the BS), but also required for beamforming or precoding in the downlink (to precisely focus the transmit energy towards the  users and to mitigate multi-user interference).
However, the fading nature of wireless channels as well as pilot contamination, i.e., channel training that may be contaminated by pilots or data transmission of users communicating in adjacent cells~\cite{Marzetta2010}, render the acquisition of accurate CSI without a significant channel-training overhead a difficult task.

Most proposed large wireless systems rely on time-division duplexing (TDD) and perform pilot-based channel training during the uplink phase~\cite{Rusek2012,larsson2014massive}.
It is, however, well known that the quality of CSI can be improved significantly by \emph{joint channel estimation and data detection} (JED), which is capable of approaching the performance of idealistic systems with perfect CSI~\cite{stoica2003space,vikalo2006efficient}.
 \rev{While a few JED algorithms have been proposed for traditional, small-scale wireless systems~\cite{stoica2003space,vikalo2006efficient,alshamary2015optimal,xu2008exact,PhamJED,prasad2015bayes,kofidis2016tensor,wen2016bayes}, their computational complexity is typically high and not much is known about their efficacy for systems with hundreds or thousands of antennas. Moreover, with the exception of the recently proposed VLSI designs in~\cite{castaneda2016}, no hardware implementations for JED have been described in the literature.}

\subsection{Contributions}
In this paper, we propose a novel, computationally efficient and near-optimal JED algorithm for large SIMO wireless systems, and we develop corresponding very-large scale integrated (VLSI) designs. 
Our contributions are summarized as follows:
\begin{itemize}
\item We use biconvex relaxation (BCR)~\cite{shah2016biconvex} to develop PRojection Onto conveX hull (\algname), a novel algorithm that achieves near-optimal JED performance at low complexity.
\item We theoretically analyze the convergence of \algname{}, which solves a biconvex problem via alternating optimization. 
\item We introduce an approximation that significantly reduces the preprocessing complexity of PrOX at virtually no loss in terms of error-rate performance.
\item We provide simulation results that showcase the robustness of PrOX to a broad range of system parameters.
\item We develop a scalable VLSI architecture for \algname{} that uses a linear array of processing elements (PEs) to achieve high-throughput at low hardware complexity. 
\item We show reference FPGA and ASIC implementation results, and compare our designs to that of the recently-reported JED implementations in \cite{castaneda2016}.
\end{itemize}
Our results demonstrate that \algname{} provides near-optimal JED at significantly lower complexity than existing reference designs.

\subsection{Related Relevant Results}

While a considerable number of algorithms and VLSI designs have been proposed for small- and large-scale multi-antenna wireless systems that separate channel estimation and data detection (see, e.g., \cite{Wu2014,yang2015fifty,castaneda2016} and the references therein), only a few of results have been proposed for JED. 
Sphere-decoding (SD) algorithms have been proposed to perform \emph{exact} maximum-likelihood (ML) JED in SIMO and MIMO systems that use a small number of time slots~\cite{stoica2003space,stoica2003joint,vikalo2006efficient,xu2008exact,alshamary2015optimal,alshamary2016efficient}.
Unfortunately, the complexity of SD methods quickly becomes prohibitive for larger dimensional problems~\cite{burg2005vlsi,studer2010vlsi}  and  approximate linear methods, which are widely used for coherent data detection in massive MIMO systems \cite{Wu2014}, cannot be used for JED (see \fref{sec:JEDdetails} for the reasons). 
Very recently, a handful of approximate JED algorithms have been proposed for large wireless systems~\cite{schenk2013noncoherent,castaneda2016,yammine2016soft,feng2017noncoherent}. 
To the best of our knowledge, reference \cite{castaneda2016} describes the only VLSI design of a JED algorithm reported in the open literature.
\rev{While this design achieves near-ML-JED performance, the algorithm relies on semidefinite relaxation (SDR), which lifts the dimensionality of the problem to the square of the number of time slots causing high hardware complexity.
In contrast to all these results, PrOX avoids lifting while achieving near-ML-JED performance and can be implemented efficiently in VLSI, even for scenarios that operate with a large number of time slots.}

\rev{Another line of related work investigates data detection algorithms that are robust to imperfect CSI~\cite{goldsmith2010robust,ghods2017optimally}. The idea is to statistically model channel estimation errors and to solve suitably adapted data detection problems. The resulting algorithms, however, (i) do not achieve the error-rate performance of JED as they are not taking into account all the received information (i.e., from training and data symbols received over multiple time slots) and (ii) do not improve the channel estimate itself---the latter is critical in TDD systems that perform beamforming in the downlink using acquired CSI in the uplink~\cite{Marzetta2010,Rusek2012,hoydis2013massive,larsson2014massive}. In contrast, we propose JED algorithms that (i) approach optimal error-rate performance as they jointly process all received information and (ii) generate improved channel estimates that can be used for beamforming.}

\subsection{Notation}
Lowercase and uppercase boldface letters stand for column vectors and matrices, respectively. For the matrix $\bA$, the Hermitian is $\bA^H$ and the $k$th row and $\ell$th column entry is $A_{k,\ell}$. For the vector $\bma$, the $k$th entry is~$a_k$.
The Euclidean norm of~$\bma$, the Frobenius norm, and the spectral norm of~$\bA$ are denoted by~$\|\bma\|_2$, $\|\bA\|_F$, and $\|\bA\|$, respectively. 
The real and imaginary parts of the vector $\bma$ are $\Re(\bma)$ and $\Im(\bma)$, respectively.

\subsection{Paper Outline}
\rev{The rest of the paper is organized as follows.} \fref{sec:system} describes the system model as well as ML-optimal JED.
\fref{sec:algo} introduces the PrOX algorithm and provides a theoretical convergence analysis. 
\fref{sec:impl} details the VLSI architecture for PrOX.
\fref{sec:results} shows error-rate performance and VLSI implementation results, and compares PrOX to existing FPGA and ASIC designs.
We conclude in \fref{sec:conclusions}. 

%%%%%%%%%%%%%%%%%%%%%%%%%%%%%%%%%%%%%%%%%%%%%%%%%%%%

\section{System Model and ML-Optimal JED}
\label{sec:system}

We now introduce the considered SIMO system model and develop the associated ML-JED problem. 

\subsection{System Model}
We study a (potentially large) SIMO wireless uplink system in which a single-antenna user transmits data over $K+1$ time slots to a BS with $B$ antennas.
We consider the standard block-fading, narrow-band\rev{\footnote{\rev{An extension to wideband systems that use OFDM~\cite{OFDM2004} is straightforward as the flat-fading model in \fref{eq:uplinkmodel} remains to be valid per active sub-carrier. An extension of our methods to channels with multi-user interference or wideband channels with inter-symbol interference is, however, not straightforward; see also Remark~\ref{rem:limitations} for more details.}}} channel model with the following input-output relation~\cite{stoica2003space,vikalo2006efficient,stoica2003joint,alshamary2015optimal}:
\begin{align} \label{eq:uplinkmodel}
\bY=\bmh\bms^H+\bN.
\end{align}
Here, the matrix $\mathbf{Y}\in\mathbb{C}^{\MR\times (K+1)}$ contains the $B$-dimensional receive vectors for the $K+1$ time slots, i.e., $\bY=[\bmy_1,\bmy_2,\ldots,\bmy_{K+1}]$ with $\bmy_k\in\complexset^B$ and $k=1,2,\ldots,K+1$,  $\bmh\in\mathbb{C}^{\MR}$ is the unknown SIMO channel vector (assumed to remain constant over the $K+1$ time slots), $\bms\in\setO^{K+1}$ contains the transmitted data symbols for all $K+1$ time slots, and $\bN\in\mathbb{C}^{\MR \times (K+1)}$ models i.i.d.\ complex circularly-symmetric Gaussian noise with variance~$\No$ per entry. 
In the remainder of the paper, we assume constant-modulus constellations, i.e., $|s|=\sigma$ for all $s\in\setO$ and a fixed $\sigma>0$. 

\begin{rem}\label{rem:limitations}
The SIMO case is relevant in many rural deployment scenarios in which only a few users are active at a time~\cite{3GPP_TS_36.211_v8.6.0}.
The assumption of having constant-modulus constellations limits our results to low-rate modulation schemes, such \rev{as} BPSK and PSK constellations. 
\rev{While the (multi-user) MIMO case and higher-order QAM modulation schemes would be of interest in more general deployment scenarios, the associated ML-JED problem is significantly more challenging to solve, and requires different, more complex, and complicated algorithms; see, e.g.,~\cite{xu2008exact,alshamary2016efficient} for more details---we are planning to address this scenario in the future.}
\end{rem}

\subsection{Joint Channel Estimation and Data Detection (JED)}
\label{sec:JEDdetails}
Let $\bmh$ be a deterministic---but unknown---channel vector with unknown prior statistics. Then, one can formulate the following ML-JED problem \cite{alshamary2015optimal}:
\begin{equation} \label{eq:JEDproblem}
\big\{\hat{\bms}^\text{JED},\hat{\bmh}\big\} = \argmin_{\bms\in\setO^{K+1},\,\bmh\in\mathbb{C}^\MR} \big\|\bY-\bmh\bms^H\big\|_F.
\end{equation}
This problem aims at simultaneously finding an estimate for the transmitted data vector $\bms$ and the channel vector $\bmh$. 
We emphasize that there is a phase ambiguity between both output vectors of ML-JED in \eqref{eq:JEDproblem}. Specifically, if $\hat\bms^\text{JED}e^{j\phi}\in\setO^{K+1}$ for $\phi\in[0,2\pi)$, then $\hat\bmh e^{j\phi}$ is also a valid solution to~\eqref{eq:JEDproblem}. \rev{In order to avoid this phase ambiguity, one can either transmit information as phase differences among the entries in the vector $\bms$, which is known as differential signaling \cite{schenk2013noncoherent}, or  fix the phase of at least one entry in the transmit vector $\bms$.
As in \cite{castaneda2016}, we set the first entry of the transmit vector to a fixed constellation point $\check{s}\in\setO$ and exploit this knowledge at the receiver.}

\begin{rem}
\rev{While this approach resembles that of conventional, pilot-based data transmission, we emphasize that ML-JED is fundamentally different. 
In traditional, pilot-based data transmission, a small number of known training symbols are used to generate channel estimates, which are then used during the detection of the data symbols. 
In contrast, ML-JED uses \emph{all} received symbols, i.e., pilots \emph{and} data symbols, to improve the quality of CSI.} 
\rev{We also note that  ML-JED as in~\eqref{eq:JEDproblem} \emph{jointly} solves for the transmitted data vector $\bms$ and the channel vector~$\bmh$; this is in contrast to JED methods  that alternate between channel estimation and data detection (see, e.g.,~\cite{PhamJED,prasad2015bayes,kofidis2016tensor,wen2016bayes}) and for which ML-JED optimality can, in general, not be guaranteed.}
\end{rem}

Since we assumed the entries in $\bms$ to be constant-modulus, the ML-JED problem in \fref{eq:JEDproblem} can be rewritten in the following, more compact form \cite{alshamary2015optimal}:
\begin{equation} \label{eq:reformulatedproblem}
\hat{\bms}^\text{JED} = \argmax_{\bms\in\setO^{K+1}}\, \|\bY\bms\|_2,
\end{equation}
and  the associated channel estimate is given by $\hat\bmh=\bY\hat\bms^\text{JED}/\|\hat\bms^\text{JED}\|^2_2$. 
We note that the optimization problem in \fref{eq:reformulatedproblem} resembles the famous MaxCut problem that is known to be NP-hard~\cite{DSteinbergThesis}. Indeed, general problems of the form~\eqref{eq:reformulatedproblem} are known to be NP-hard with respect to randomized reductions, even when only approximate solutions are sought \cite{ajtai1998shortest}.
Nevertheless, for a small number of time slots $K+1$, the problem in \eqref{eq:reformulatedproblem} can be solved exactly and at low average complexity using sphere decoding (SD) methods~\cite{alshamary2015optimal}. For a large number of time slots, however, SD is known to entail prohibitively high complexity~\cite{seethaler2011complexity}.
Furthermore, linear approximations are useless for JED, since the entries of $\bms$ would grow without bound when relaxing the finite-alphabet constraint $\bms\in\setO^{K+1}$ to the complex numbers $\bms\in\mathbb{C}^{K+1}$.
As a consequence, the design of non-linear and low-complexity algorithms is necessary to enable JED for practical SIMO systems that use a large number of time slots. 

%%%%%%%%%%%%%%%%%%%%%%%%%%%%%%%%%%%%%%%%%%%%%%%%%%%%

\section{\algname: PRojection Onto conveX hull}
\label{sec:algo}

We now develop a novel algorithm to approximately solve the ML-JED problem in \fref{eq:reformulatedproblem} using biconvex relaxation (BCR)~\cite{shah2016biconvex}, a recent framework to solve large semidefinite programs in computer vision. The resulting algorithm is referred to as PRojection Onto conveX hull (\algname), requires low computational complexity, and achieves near-ML-JED performance. 

\subsection{Biconvex Relaxation (BCR) of the ML-JED Problem}
We start from \fref{eq:reformulatedproblem} and include a regularization term that forces the transmit vector $\bms\in\setO^{K+1}$ to be close to a copy $\bmq\in\C^{K+1}$ that is relaxed to the complex numbers as follows:
\begin{equation} \label{eq:reformulatedproblemalpha}
\underset{\bms\in\setO^{K+1},\bmq\in\C^{K+1}}{\mathrm{minimize}} \, -\|\bY\bmq\|_2^2 + \alpha\|\bmq-\bms\|_2^2,
\end{equation}
where $\alpha>0$ is a suitably chosen regularization parameter.
To ensure that the problem in \eqref{eq:reformulatedproblemalpha} is convex in the vector $\bmq$, the parameter $\alpha$ must be larger than the maximum eigenvalue of the Gram matrix $\bG=\bY^H \bY$, i.e., $\alpha>\|\bY^H \bY\|$.

We next relax the finite-alphabet constraint to the convex hull $\mathcal{C}_\setO$ of the set $\setO$ given by~\cite{Wu2016ocd}
\begin{equation*}
\mathcal{C}_\setO = \left\lbrace
 \sum_{i=1}^{|\setO|} \alpha_i s_i \mid (\alpha_i \in \R^+, \forall i )
 \wedge
 \sum_{i=1}^{|\setO|} \alpha_i = 1
\right\rbrace\!,
\end{equation*}
with $s_i$, $i=1,\ldots,|\setO|$. 
For example, the convex hull for BPSK constellations $\mathcal{C}_{\text{BPSK}}$ is the line along the real axis in $[-1,+1]$; the convex hull $\mathcal{C}_{\text{QPSK}}$ for QPSK is the square with the four constellation points as corners. 
By relaxing $\setO$ to the convex hull $\setC_\setO$, we arrive at the following relaxed ML-JED problem:
\begin{equation*} 
\underset{\bms\in\mathcal{C}^{K+1}_\setO,\bmq\in\C^{K+1}}{\mathrm{minimize}} \, -\|\bY\bmq\|_2^2 + \alpha\|\bmq-\bms\|_2^2.
\end{equation*}

For the above problem, it is likely to obtain a solution that is inside the convex hull. We are, however, interested in finding solutions that lie at the boundary of the convex hull~$\setC_\setO$ as the original constellation points in $\setO$ lie at the boundary. 
To this end, we include a norm constraint that promotes large values in $\bms$, with the goal of pushing the entries of $\bms$ towards the boundary of the convex hull. This leads to the final BCR formulation of the ML-JED problem:
\begin{align} \label{eq:BCRform}
\hat{\bms}^\text{BCR} = \argmin_{\bms\in\mathcal{C}^{K+1}_\setO,\bmq\in\C^{K+1}} -\|\bY\bmq\|_2^2 + \alpha\|\bmq-\bms\|_2^2 - \beta\|\bms\|_2^2.
\end{align}
Here, $\beta$ is a suitably chosen algorithm parameter that satisfies \rev{$\beta<\alpha$}, which ensures that the optimization problem is \emph{biconvex} in the vectors $\bms$ and $\bmq$. In words, for a fixed vector~$\bmq$, the problem in \fref{eq:BCRform} is convex in~$\bms$, and vice versa. 

\subsection{Alternating Optimization}
We solve the BCR problem in \eqref{eq:BCRform} using alternating minimization, i.e., we keep one of the vectors fixed while solving for the other. The resulting iterative procedure is given by
\begin{align}
\bmq^{(t)} &= \argmin_{\bmq \in\mathbb{C}^{K+1}}  \,-\|\bY\bmq\|_2^2 + \alpha\|\bmq-\bms^{(t-1)}\|_2^2 \label{minq}\\
\bms^{(t)} & = \argmin_{\bms\in\mathcal{C}^{K+1}_\setO} \,\alpha\|\bmq^{(t)}-\bms\|_2^2 -\beta\|\bms\|_2^2, \label{mins}
\end{align}
where $t=1,2,\ldots,t_\text{max}$ is the iteration counter. 

We initialize the above algorithm with $\bms^{(0)}=\check{s}(G_{1,1})^{-1}\bmg^c_1$, where~$\bmg^c_1$ is the first column of $\bG$, and $\check{s}\in\setO$ is the fixed symbol transmitted in the first time slot and known at the receiver.

Since the problem in \eqref{eq:BCRform} is biconvex in $\bms$ and $\bmq$, both of the above steps are convex and can be solved optimally. Furthermore, each of these optimization problems turns out to have a closed form solution. Concretely, we obtain the following two-step PrOX algorithm:
\begin{align}
\bmq^{(t)} &= (\bI - \alpha^{-1}\bG)^{-1} \bms^{(t-1)} \label{eq:origpush1} \\
\bms^{(t)} & = \textrm{prox}_{\mathcal{C}^{K+1}_\setO} (\theta \bmq^{(t)}), \label{eq:origpush2}
\end{align}
where $\theta=\frac{\alpha}{\alpha-\beta}>0$, $\bI$ is the identity matrix, and the proximal operator~\cite{nealboyd2013proximal} is defined as
\begin{align} \label{eq:proxy}
\textrm{prox}_{\mathcal{C}^{K+1}_\setO}(\bmv)=\argmin_{\bms\in\mathcal{C}^{K+1}_\setO} \| \bmv - \bms \|^2_2.
\end{align} 
\rev{This proximal operator is the element-wise orthogonal projection of the vector $\bmv=\theta\bmq^{(t)}$ onto the convex hull~$\mathcal{C}_\setO$ around the constellation set~$\setO$.}
For BPSK, the proximal operator in~\eqref{eq:proxy} is given by  $\textrm{prox}_{\mathcal{C}_\text{BPSK}} ( v_b ) = \max \left\lbrace \min \left\lbrace \Re(v_b), +1 \right\rbrace, -1 \right\rbrace$, i.e., the projection onto the real line in $[-1,+1]$; for QPSK, \eqref{eq:proxy} corresponds to applying the same operation, independently, to both $\Re(v_b)$ and~$\Im(v_b)$.
In the case of \eqref{eq:origpush2}, we have $\bmv=\theta \bmq^{(t)}$ with \rev{$\theta>0$} and hence, the proximal operator in \eqref{eq:proxy} projects a \rev{scaled} version of~$\bmq^{(t)}$ onto the convex hull of the constellation set~$\mathcal{C}_\setO$---this is why we dub our algorithm \emph{PRojection Onto conveX hull} (\algname{}). 

\subsection{Convergence Theory of PrOX}

We now show that the PrOX algorithm is well behaved, even though it solves a biconvex problem using alternating optimization.
More specifically, we begin by establishing that the iterates $\bmq^{(t)}$ and  $\bms^{(t)}$ generated by PrOX converge to stationary points of the objective function in~\eqref{eq:BCRform}, provided that the algorithm parameters $\alpha$ and $\beta$ are chosen appropriately.  We then show that these stationary points lie on the boundary of the convex hull $\setC_\setO^{K+1},$ which implies that our biconvex relaxation is reasonably tight.
In what follows, we denote the objective function of PrOX in~\eqref{eq:BCRform} as
\begin{align*}
f(\bmq,\bms) = \left\{\begin{array}{l}
  -\frac{1}{2}\|\bY\bmq\|_2^2 + \frac{\alpha}{2}\|\bmq-\bms\|_2^2 - \frac{\beta}{2}\|\bms\|_2^2, \text{ if } \bms\in\setC_\setO^{K+1}\\[0.2cm]
   \infty, \text{ otherwise.} 
  \end{array}\right.
\end{align*} 
\rev{Note that the factors of $\frac{1}{2}$ do not affect the solution of PrOX.}

At an optimal solution, the sub-gradients of $f$ with respect to both $\bms$ and $\bmq$ should be zero.  Because $\bms^{(t)}$ is computed by minimizing $f$ with respect to $\bms^{(t)},$ we already know that the sub-gradient of $f$ with respect to $\bms^{(t)}$ contains zero.
We can thus measure the optimality of the iterates by examining only $\nabla_\bmq f,$ the gradient of $f$ with respect to $\bmq.$  We now show that this gradient vanishes for a large number of iterations $t$; the proof is given in \fref{app:convergence1}.
\begin{thm} \label{thm:convergence1}
\rev{Suppose that the parameters $\alpha$ and $\beta$ satisfy $\alpha>\|\bG\|$ and $\alpha>\beta$}. Then, the PrOX algorithm in \fref{eq:origpush1} and~\fref{eq:origpush2} converges in the gradient sense, i.e.,
$$\lim_{t\to\infty} \nabla_{\bmq}f(\bmq^{(t)},\bms^{(t)})=0.$$
Furthermore, any limit point of the sequence of iterates is a stationary point.
\end{thm}

The PrOX algorithm is founded on the idea of replacing the discrete constellation $\setO$ with its convex hull $\setC_\setO.$  This results in a biconvex problem that is easily solved without resorting to expensive discrete programming or lifting-based methods.  However, in using a biconvex relaxation, there is a risk of finding a minimizer that lies somewhere in the interior of the convex hull $\setC_\setO,$ far away from any of the constellation points.  We now provide conditions for which this situation does not happen. The following theorem shows that minimizers of \eqref{eq:BCRform} always lie on the boundary of the convex hull $\setC_\setO$; a proof is given in \fref{app:convergence2}.
\begin{thm} \label{thm:convergence2}
Let the conditions of \fref{thm:convergence1} hold and further assume $\alpha > \beta > 0$. Then, any non-zero stationary point of \eqref{eq:BCRform} lies along the boundary of the set $\setC_\setO.$
\end{thm}

\begin{rem}
\fref{thm:convergence1} and \fref{thm:convergence2} do not guarantee PrOX to find a \emph{global} minimizer to the ML-JED problem \fref{eq:reformulatedproblem}, but rather stationary (or saddle) points  that lie on the boundary of the convex hull---although we often observe global minimizers in practice (cf.~\fref{sec:errorrate}). 
The development of stronger results that provide conditions for which PrOX finds the optimal solution is challenging and left for future work.
\end{rem}

\subsection{Simplifying the Preprocessing Stage}
\label{sec:preprocessingtrick}
For a large number of time slots $K+1$, computing the matrix inverse $\widehat\bG=(\bI-\alpha^{-1}\bG)^{-1}$ in \fref{eq:origpush1} results in high preprocessing complexity. We now propose an approximate method that substantially reduces the preprocessing complexity for PrOX at a negligible loss in terms of error-rate performance. 

Let $\|\bG\|<\alpha$ for any consistent matrix norm $\|\cdot\|$. Then, we have the following Neumann series expansion~\cite{Stewart1998}:
\begin{align} \label{eq:neumannseries}
(\bI-\alpha^{-1}\bG)^{-1} = \sum_{k=0}^\infty (\alpha^{-1}\bG)^k.
\end{align}
As put forward in \cite{Wu2014} for data detection in massive MIMO systems, one can approximate a matrix inverse by truncating the series expansion to the first two terms, which is 
\begin{align} \label{eq:approximationpreprocessing}
(\bI-\alpha^{-1}\bG)^{-1}  \approx \bI + \alpha^{-1}\bG.
\end{align}
\rev{Evidently, the expression on the right-hand side does not require the computation of a matrix inverse, which significantly reduces the preprocessing complexity. 
More specifically, the matrix to be inverted in the left-hand side of \fref{eq:approximationpreprocessing} is of dimension $(K+1)\times(K+1)$. The complexity (in terms of the number of real-valued multiplications) of an explicit matrix inversion using, e.g., the Cholesky factorization scales with $(K+1)^3$ and exhibits stringent data dependencies when implemented in VLSI;  see~\cite{Wu2014} for more details on the computational complexity of Cholesky-based matrix inversion and a VLSI design. Hence, for a large number of time slots~$K$, the approximate preprocessing method in \fref{eq:approximationpreprocessing} yields significant savings in terms of hardware complexity.}

We have the following result that bounds the error of the approximation in \fref{eq:approximationpreprocessing}; the proof is given in \fref{app:preprocessingapprox}.
\begin{thm} \label{thm:preprocessingapprox}
Let  $\alpha>\|\bG\|$. Then, the error of the approximation in \fref{eq:approximationpreprocessing} is bounded by 
\begin{align*}
\|(\bI-\alpha^{-1}\bG)^{-1} - (\bI + \alpha^{-1}\bG) \| \leq \frac{\|\alpha^{-1}\bG\|^2}{1-\|\alpha^{-1}\bG\|}.
\end{align*}
\end{thm}

This result implies that if $\|\bG\|$ is smaller than $\alpha$, then the approximation error will be small. In other words, the approximation error will depend on the parameter $\alpha$, which we can tune in practice for optimal performance. In \fref{sec:errorrate}, we will show that PrOX with this approximation results in excellent error-rate performance while significantly reducing the complexity compared to using the original matrix $\widehat\bG$.

\subsection{Hardware-Friendly Variant of \algname}
As a last step, we now modify the PrOX algorithm in \eqref{eq:origpush1} and \eqref{eq:origpush2}  to make it more hardware friendly. 
Since the BS is assumed to know the first entry of $\bms$, there is no need to apply the algorithm to this particular entry. Consequently, we force the entry~$s_1^{(t)}$ to be $\check{s}$ at the end of each iteration.

The matrix~$\widehat\bG$ exhibits, in general, a large dynamic range for different system configurations and channel realizations.
To facilitate fixed-point design, we divide all of its elements by a constant $\gamma > 0$, so that the entries of the resulting matrix are close to one in absolute value. 
This scaling procedure requires us to introduce a new vector $\tilde{\bmq}=\gamma^{-1}\bmq$, resulting in the following modified PrOX algorithm:
\begin{align}
\tilde{\bmq}^{(t)}&=\gamma^{-1}\bmq^{(t)} = \gamma^{-1} \widehat\bG \bms^{(t-1)} \label{eq:iterpush1} \\
\bms^{(t)} &
= \textrm{prox}_{\mathcal{C}^{K+1}_\setO} (\theta \gamma \tilde{\bmq}^{(t)}) 
= \textrm{prox}_{\mathcal{C}^{K+1}_\setO} (\varrho \tilde{\bmq}^{(t)}),  \label{eq:iterpush2} 
\end{align}
where $\varrho = \theta \gamma$.
With these two modifications, we arrive at the hardware-friendly version of \algname{} summarized as follows: 

\begin{oframed}
\vspace{-0.2cm}
\begin{alg}[Hardware-Friendly PrOX]\label{alg:PrOX}  
Fix the algorithm parameters $\varrho>0$ and $\alpha>\|\bG\|$. Precompute the matrix~$\widehat\bG$ using one of the following expressions:
\begin{align} 
\widehat{\bG} & = \gamma^{-1}(\bI - \alpha^{-1}\bG)^{-1}  \label{eq:regularizedmatrix1} \\
\widehat{\bG} & = \gamma^{-1}(\bI + \alpha^{-1}\bG), \label{eq:regularizedmatrix2}
\end{align} 
and initialize $\bms^{(0)}=\check{s}(G_{1,1})^{-1}\bmg^c_1$.
Then, for every iteration $t=1,2,\ldots,t_\text{max},$ compute the following steps:
\begin{align}
\tilde{\bmq}^{(t)}&= \widehat{\bG}\bms^{(t-1)} \label{eq:step1}\\
\bms^{(t)} & = \mathrm{prox}_{\mathcal{C}^{K+1}_\setO}(\varrho \tilde{\bmq}^{(t)} ) \label{eq:step2} \\
s^{(t)}_1 & =\check{s}  \label{eq:step3} 
\end{align}
At the end of iteration $t_\text{max}$, compute hard-output estimates of the transmitted signals as follows:
\begin{align}
\hat{s}_k = \argmin_{s\in\setO} |s_k^{(t_\text{max})}-s|, \quad k=1,2,\ldots,K+1.  \label{eq:step4} 
\end{align}
\end{alg}
\vspace{-0.3cm}
\end{oframed}
In what follows, we will use PrOX to refer to \fref{alg:PrOX} with exact preprocessing as in \fref{eq:regularizedmatrix1} and approximate PrOX~(APrOX) to refer to \fref{alg:PrOX} with approximate preprocessing as in \fref{eq:regularizedmatrix2}. 
We also note that $\alpha>0$ and $\varrho>0$ are algorithm parameters that can be tuned via numerical simulations to  improve the error-rate performance. 

%%%%%%%%%%%%%%%%%%%%%%%%%%%%%%%%%%%%%%%%%%%%%%%%%%%%

%\setlength{\textfloatsep}{10pt}% Remove \textfloatsep
\begin{figure*}[tp]
\centering
%\vspace{-0.1cm}
\includegraphics[width=0.99\textwidth]{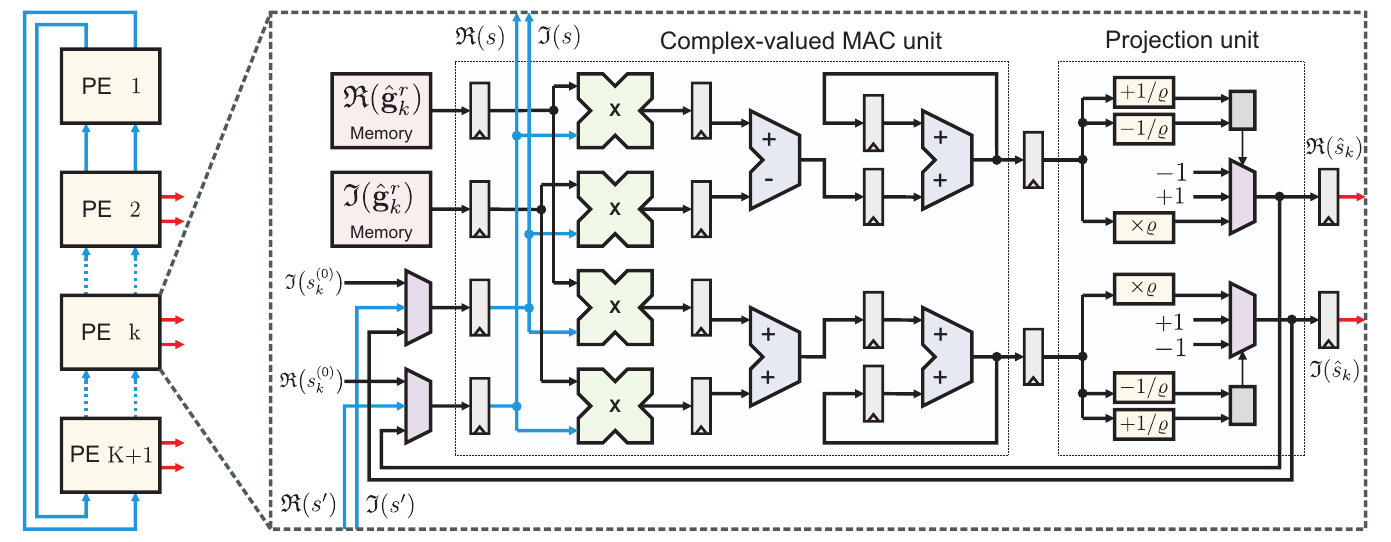}
%\vspace{-0.1cm} 
\caption{VLSI architecture of \algname{}. Left: linear array of $K+1$  processing elements~(PEs), which enables us to achieve high throughput at low complexity\rev{. Right:} architecture details of the $k$th PE, which mainly consists of a complex-valued multiply-accumulate (MAC) unit and a projection unit.}
\label{fig:arch}
\vspace{-0.13cm} 
\end{figure*}

\section{VLSI Architecture}
\label{sec:impl}

We now propose a VLSI architecture for PrOX, which exhibits high regularity and enables us to achieve high throughput at low hardware complexity. 

\subsection{Architecture Overview}
\rev{\fref{fig:arch} shows the VLSI architecture for \algname{} in a system that uses QPSK\footnote{\rev{For BPSK modulation, the VLSI architecture shown in \fref{fig:arch} can be simplified as data-path elements used to compute $\Im\left(\bms\right)$ can be removed.}} modulation. At a high level, our architecture consists of a linear array of $N=K+1$ processing elements~(PEs), each one dedicated to computing an entry of the vectors~$\bms$ and~$\tilde{\bmq}$.}
\rev{To execute \fref{alg:PrOX}, each PE \rev{has} three key components: (i) a $\widehat{\bG}$-matrix memory, (ii) a complex-valued multiply-accumulate (MAC) unit, and (iii) a projection unit (see the right side of \fref{fig:arch}).}
\rev{The $\widehat{\bG}$-matrix memory comprises two memories to store the real and imaginary parts of $\hat{\bmg}_k^r$,  i.e., the $k$th row of the matrix~$\widehat{\bG}$.} 
\rev{We assume that $\widehat{\bG}$, which is the result of either~\fref{eq:regularizedmatrix1}~(for \algname{}) or \fref{eq:regularizedmatrix2}~(for \apalgname{}), was computed and loaded in the memories during a preprocessing step.}
\rev{The MAC unit of the $k$th PE is used to sequentially compute the $k$th row of the matrix-vector product (MVP) in Step~\fref{eq:step1} of \fref{alg:PrOX}, resulting in~$\tilde{q}^{(t)}_k$ (see~\fref{sec:mvpoperation} for the details).}
\rev{Finally, the projection unit of the $k$th PE computes the $k$th entry of Step~\fref{eq:step2} of \fref{alg:PrOX}, i.e., it uses $\tilde{q}^{(t)}_k$ to acquire $s^{(t)}_k$.}
\rev{Since $s^{(t)}_1=\check{s}$, the first PE does not need the previous three key components. Instead, PE~1 only contains two multiplexers and flip-flops that store the predefined constellation point $\check{s}$ in order to implement Step \fref{eq:step3} of \fref{alg:PrOX}.}
\rev{The implementation of all other PEs is identical and uses the aforementioned three key components. For these PEs, the hard-output estimates in Step~\fref{eq:step4} are extracted directly from the sign bits at the outputs of the projection units.}

\subsection{Matrix-Vector Product (MVP)}
\label{sec:mvpoperation}
\begin{figure*}[tp]
\centering
%\vspace{-0.1cm}
\subfigure[Conventional column-by-column matrix-vector product]{\includegraphics[width=1.6\columnwidth]{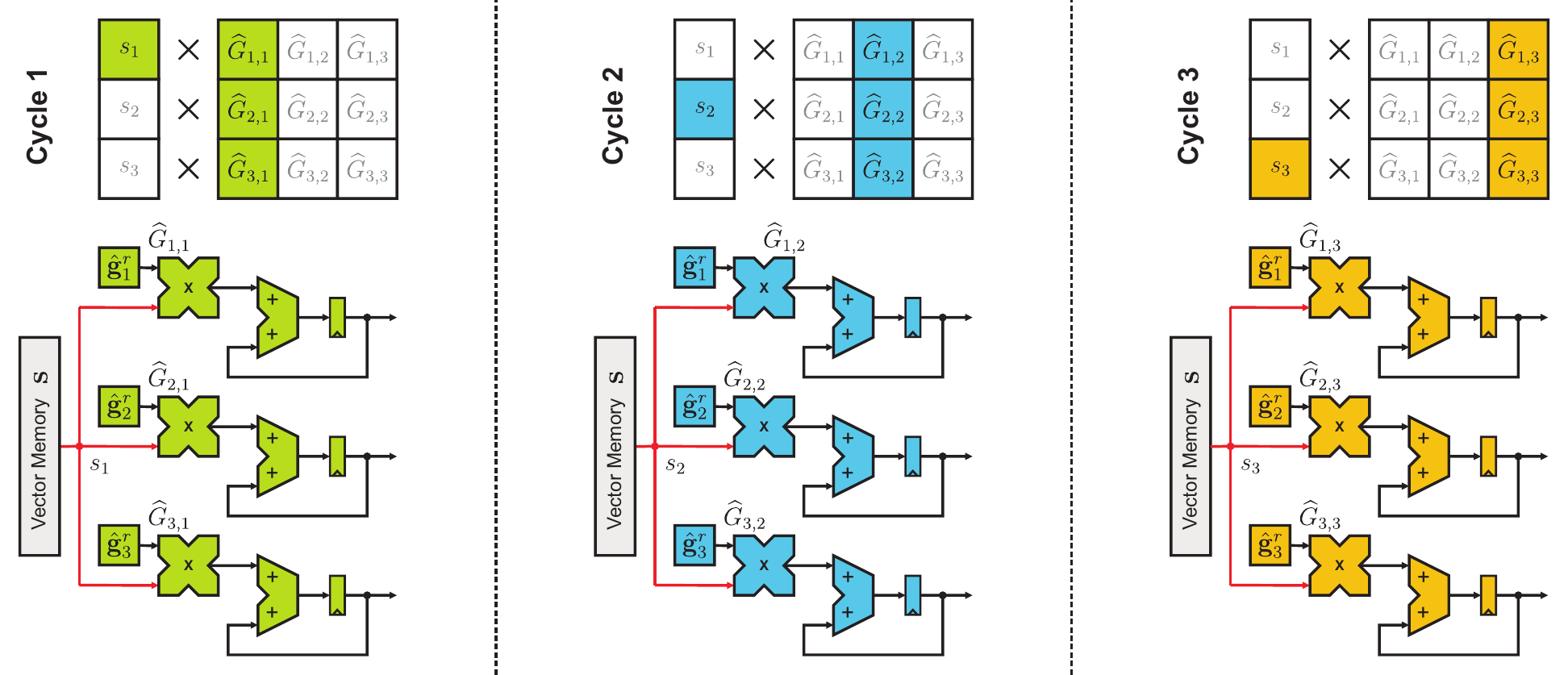}\label{fig:MVPoperation1}}
\subfigure[Proposed input-cyclic matrix-vector product for PrOX]{\includegraphics[width=1.6\columnwidth]{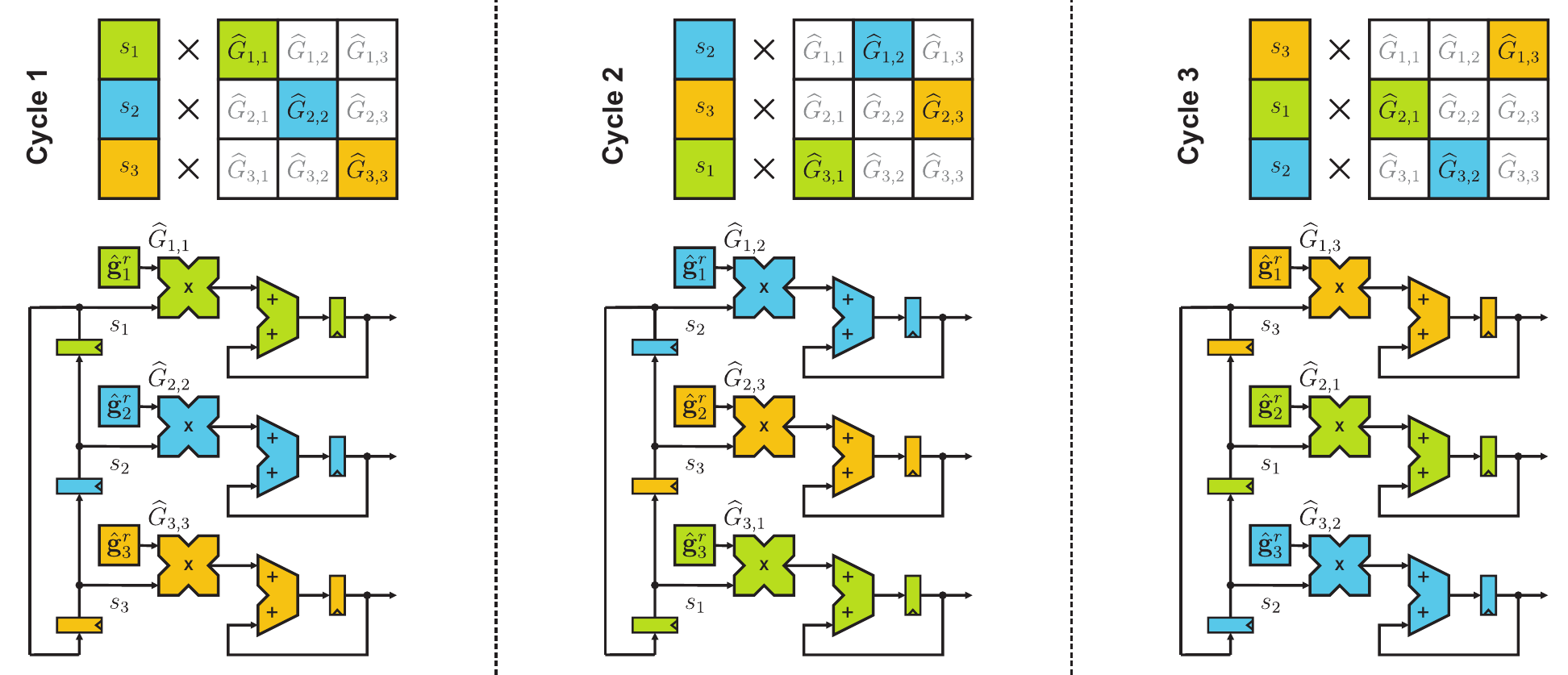}\label{fig:MVPoperation2}}
%\vspace{-0.1cm} 
\caption{Processing details of the matrix-vector product (MVP) operation: (a) conventional approach that proceeds on a column-by-column basis and leads to high memory fan-out; (b) proposed input-cyclic method that reduces fan-out.}
\label{fig:MVPoperation}
\end{figure*}

\rev{Before explaining the operation details of the proposed VLSI architecture, we focus on the main computation of PrOX: the MVP operation in Step~\fref{eq:step1} of \fref{alg:PrOX}. A straightforward way for computing $\tilde\bmq^{(t)} = \widehat{\bG}\bms^{(t-1)}$ using a linear array of PEs is depicted in \fref{fig:MVPoperation1} for $N=K+1=3$.} This architecture computes~$\tilde\bmq^{(t)}$ sequentially and on a column-by-column basis, \rev{i.e., it evaluates $\tilde\bmq^{(t)} = \sum_{k=1}^{N}\hat{\bmg}^c_k s^{(t-1)}_k$ in a sequential manner over $k=1,\ldots,N$ clock cycles. Here,~$\hat{\bmg}^c_k$ represents the $k$th column of~$\widehat{\bG}$. While such an approach is conceptually simple, it requires a centralized vector memory which suffers from a high fan-out at its output (highlighted with red color in \fref{fig:MVPoperation1}). More concretely, the vector memory output must be connected to $N=K+1$ MAC units, eventually becoming the critical path for systems with a large number of time slots $K+1$.}
\rev{While wire pipelining at the output of the vector memory could be used to mitigate the fan-out, we propose an alternative solution that avoids this issue altogether.}

Consider the MVP architecture depicted in \fref{fig:MVPoperation2}. \rev{In this architecture, we replace the centralized vector memory by a set of registers: each register is associated with a PE and contains an entry of the vector $\bms^{(t-1)}$. Furthermore, the registers are chained together in a cyclic manner, forming a shift register that enables all entries of~$\bms^{(t-1)}$ to reach all the PEs in a round-robin fashion. 
Then, each PE only needs to access the correct entry of~$\hat\bmg^r_k$ to ensure that the accumulator of their MAC unit contains the correct MVP result after $N$ clock cycles. By following this approach, each register only drives the next register in the cycle and one of the MAC unit inputs, regardless of $N$, hereby completely avoiding the fanout issue. We henceforth call this architecture \emph{input-cyclic MVP}, which builds the core component of the PrOX architecture shown in \fref{fig:arch} and detailed in the next subsection.}

\rev{We note that the reading pattern of the~$\widehat\bG$-matrix memory for each PE corresponds to simply accessing the entries of $\hat\bmg^r_k$ in order. However, in contrast to the conventional architecture in \fref{fig:MVPoperation1}, reading can start from any element and wraps around after reaching the end of $\hat\bmg^r_k$. Instead of using address generation logic required to implement this behavior, we store a cyclically-permuted version of $\hat\bmg^r_k$ in the $\widehat{\bG}$-matrix memory of each PE. By doing so, each PE simply needs to access its $\widehat{\bG}$-matrix memory in regular order starting from the first address, as it would be done in the conventional architecture depicted in \fref{fig:MVPoperation1}.}
\subsection{Operation Details of the PrOX Architecture}
\label{sec:architectureoperation}
\rev{To implement the input-cyclic MVP architecture, the entries of the~$\widehat{\bG}$ matrix for the $k$th PE are stored in the following way: The first address of the PE memory contains $\widehat{G}_{k,k}$; the second address contains $\widehat{G}_{k,k+1}$, and so on. For example, in the case shown in \fref{fig:MVPoperation2}, the second PE would have its $\widehat{\bG}$-matrix entries organized as follows: $\{ \widehat{G}_{2,2}, \widehat{G}_{2,3}, \widehat{G}_{2,1}\} $.}

\rev{In the first clock cycle, the $k$th PE has access to both $\widehat{G}_{k,k}$ and $s^{(t-1)}_k$, so it can compute the $\widehat{G}_{k,k} s^{(t-1)}_k$ product, and store the result in its accumulator. At the same time, this PE passes its current $\bms^{(t-1)}$ entry to the subsequent PE in the chain, i.e., the $(k-1)$th PE. Passing the current $\bms^{(t-1)}$ entry to the next PE will be performed in all PEs during all the clock cycles of the MVP computation.
An example of this first clock cycle is shown in \fref{fig:MVPoperation2}, where the second PE computes $\widehat{G}_{2,2} s_2$ and passes the $s_2$ value to the first PE, so that the first PE has access to~$s_2$ in the second clock cycle.}

\rev{In the second clock cycle, the $k$th PE receives $s^{(t-1)}_{k+1}$ from the $(k+1)$th PE, the previous PE in the array. As a result, the $k$th PE can compute $\widehat{G}_{k,k+1} s^{(t-1)}_{k+1}$ and accumulate it with the previous product.
For example, in the second clock cycle of \fref{fig:MVPoperation2}, the second PE receives $s_3$ from the third PE and is hence able to compute $\widehat{G}_{2,3} s_3$ and add it with $\widehat{G}_{2,2} s_2$.}

\rev{In the third clock cycle, the $k$th PE receives $s^{(t-1)}_{k+2}$ from the $(k+1)$th PE. Note that, while the $s^{(t-1)}_{k+2}$ value was initially in the $(k+2)$th PE, the $(k+2)$th PE previously passed this value to the $(k+1)$th PE. Then, the $k$th PE has the necessary operands to continue executing its complex-valued MAC operation.
To clarify this aspect, let us examine the third clock cycle in \fref{fig:MVPoperation2}: the second PE has access to $\widehat{G}_{2,1}$ and $s_1$ to finish its computation. While the second PE received $s_1$ from the third PE, the third PE received $s_1$ from the first PE in the previous clock cycle.}

\rev{This example illustrates how the exchange of values between PEs enables each entry of $\bms^{(t-1)}$ to circulate through the linear array of PEs. In the case of the example in \fref{fig:MVPoperation2}, a complete circulation of all values in the shift registers takes three clock cycles.}
%, 
%
\rev{In general, after $K+1$ clock cycles, all PEs have had access to all the entries of $\bms^{(t-1)}$ and used them to compute their respective entry of~$\tilde{\bmq}^{(t)}$. During this procedure, the first PE (which is implemented differently than the other PEs) executes the same procedure: it passes $\check{s}$ to the $(K+1)$th PE during the first clock cycle, and in the subsequent clock cycles sends the $\bms^{(t-1)}$ entry received from the second PE to the $(K+1)$th PE.}
As the MAC units contain three pipeline stages, two clock cycles are required to flush the pipeline. Hence, the MVP operation in Step \fref{eq:step1} of \fref{alg:PrOX} is computed in only $K+3$ clock  cycles.

\rev{We now describe the operation of the projection unit shown on the right side of \fref{fig:arch}, which implements $\bms^{(t)} = \mathrm{prox}_{\mathcal{C}^{K+1}_\setO}(\varrho \tilde{\bmq}^{(t)} )$.}
\rev{For QPSK, the projection unit of the $k$th PE consists of two identical modules: one module takes $\bar{q}=\Re(\tilde{q}_k^{(t)})$ as its input, while the other module takes $\bar{q}=\Im(\tilde{q}_k^{(t)})$ as its input. For BPSK, only the module with an input of $\bar{q}=\Re(\tilde{q}_k^{(t)})$ is required.}
\rev{The function of each module is to compute $\varrho \bar{q}$ and to clip its value in case it exceeds an absolute value of $1$. In other words, each module determines if $\varrho \bar{q}$ is smaller than $-1$, larger than $+1$, or in-between these numbers, and outputs $-1$, $+1$, or $\varrho \bar{q}$, respectively.}
\rev{As $\varrho > 0$, to determine if $\varrho \bar{q} \geq +1$, one can also check $\bar{q} \geq +1/\varrho$ or, equivalently, if $\bar{q}-1/\varrho \geq 0$, which can be extracted from the sign bit of $\bar{q}-1/\varrho$. Analogously, to determine if $\varrho \bar{q} < -1$, one can also check $\bar{q}+1/\varrho < 0$, which can be extracted from the sign bit of $\bar{q}+1/\varrho$.}
\rev{As a result, each module takes its input $\bar{q}$ and computes $\bar{q}-1/\varrho$ and $\bar{q}+1/\varrho$. Using the sign bits of these two results, the module can select if the output will be $-1$, $+1$, or $\varrho \bar{q}$.}
\rev{The quantity $\varrho \bar{q}$ is computed in parallel with the calculation of $\bar{q}+1/\varrho$ and $\bar{q}-1/\varrho$, which reduces the critical path.}
By restricting the parameter $\varrho$ to be a power of two, the multiplication can be carried out with inexpensive arithmetic shifts.
\rev{The projection unit uses one clock cycle to complete its operation, but it can start only after the $K+3$ clock cycles used by the complex MAC unit. After the projection unit has finished, the new $s^{(t)}_k$ will be available at the inputs of the complex-valued MAC unit of the $k$th PE, ready to start a new iteration.}
Consequently, each \algname{} iteration requires a total number of $K+4$ clock cycles.

\subsection{Implementation Details}
We now provide the remaining implementation details, including the fixed-point parameters as well as memory considerations for our FPGA and ASIC designs.

\subsubsection{Fixed-Point Parameters}
To minimize area and power consumption, and to maximize the throughput, we deploy fixed-point arithmetic in our design.
Specifically, our architecture uses $6$\,bit signed fixed-point values for representing the $\bms^{(t)}$ entries, with $3$ fraction bits. For the elements of $\widehat{\bG}$, $12$\,bit signed fixed-point values are used, with $11$ fraction bits. \rev{While $18$ bit values are generated at the outputs of the multipliers in the complex-valued MAC unit, only $15$ bits are used in the subsequent adders and accumulators (of which $11$ are fraction bits). The adder and subtractor that receive the outputs from the multipliers do not saturate, but rather wrap-around which reduces area and delay. In contrast, the accumulator adders saturate. The outputs of the complex-valued MAC unit are also $15$\,bit signed fixed-point values with $11$ fraction bits. In the projection unit, the $-1/\varrho$ and $+1/\varrho$ quantities are represented using $12$\,bit signed fixed-point numbers with $11$ fraction bits, and are added to the outputs of the complex MAC unit using $15$\,bit saturating adders. The outputs of these adders are used to determine the output of the projection unit. Furthermore, for all the considered system configurations, $\varrho > 1$. As $\varrho$ is restricted to be a power of two, a multiplication with $\varrho$ is implemented by an arithmetic left shift. The number of shifts is encoded with a $4$\,bit number. Only the $6$ most significant bits of the shifted value are taken into the multiplexer of the projection unit, as the output of this multiplexer will become the next iterate~$\bms^{(t+1)}$. We note that these fixed-point parameters are sufficient to achieve near-floating-point performance; see Figures~\ref{fig:16x17ul} and \ref{fig:16x17dl} in \fref{sec:errorrate} for corresponding symbol error-rate performance results.}

\subsubsection{Memory Considerations}
For the FPGA designs, the $\widehat{\bG}$-matrix memory is implemented using look-up tables (LUTs) as distributed RAM, i.e., no block RAMs have been used.
For the ASIC designs, the $\widehat{\bG}$-matrix memory is implemented using latch arrays as described in \cite{meinerzhagen2010towards}; these memories are built from standard cells and simplify automated design synthesis, without a significant area penalty  compared to SRAM macrocells. 

 \section{Results}
\label{sec:results}

We now show error-rate performance results as well as FPGA and ASIC implementation results. We also compare our design to the only other existing JED design reported \rev{recently} in~\cite{castaneda2016}. 

\begin{figure*}[tp]
\centering
\subfigure[BPSK, Rayleigh fading]{\includegraphics[width=0.325\textwidth]{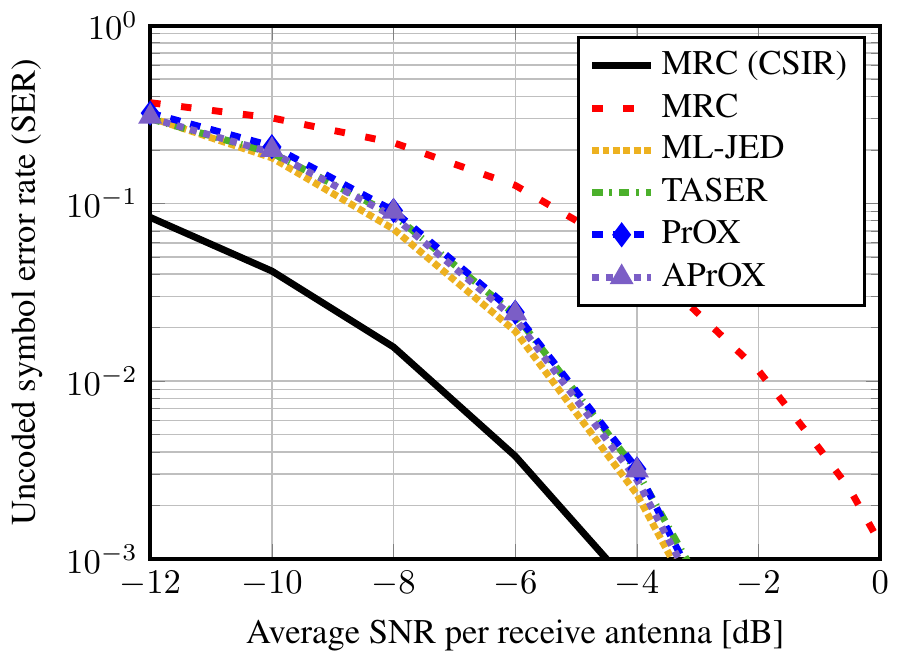}\label{fig:16x17bpskul_er}}
\subfigure[QPSK, Rayleigh fading]{\includegraphics[width=0.325\textwidth]{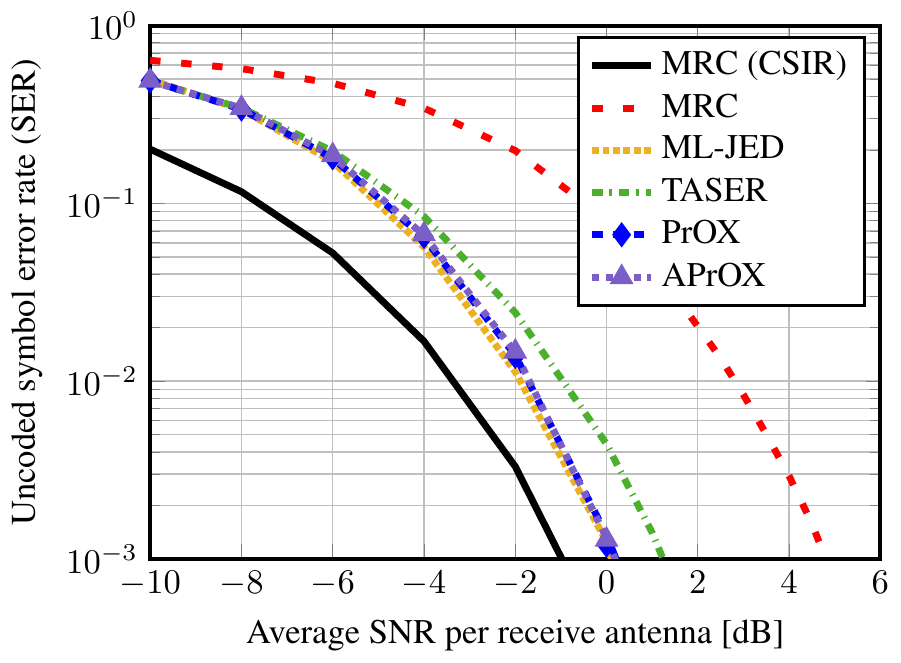}\label{fig:16x17qpskul_er}}
\subfigure[QPSK, LoS]{\includegraphics[width=0.325\textwidth]{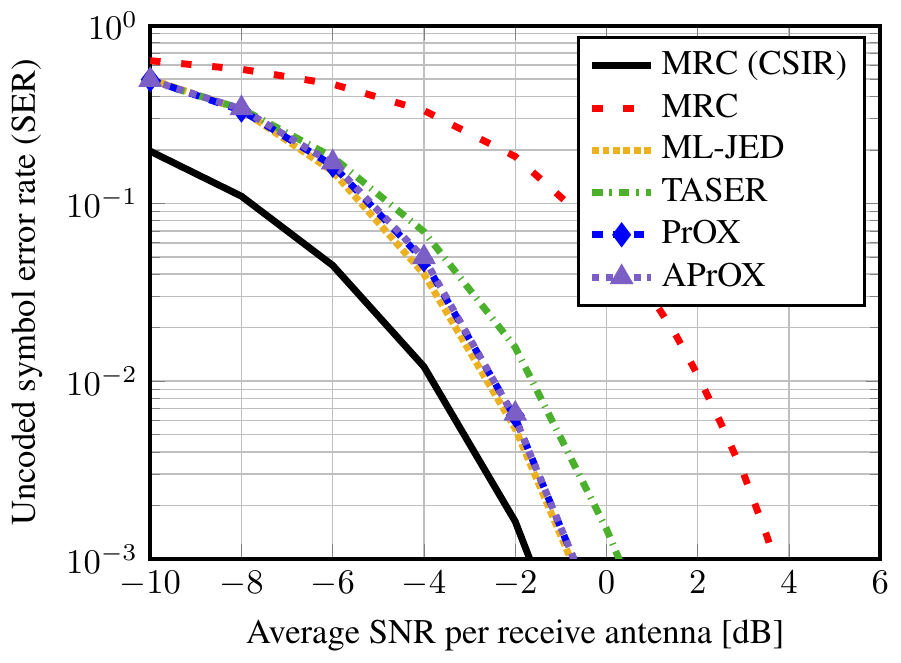}\label{fig:16x17qpsklosul_er}}
\caption{Uncoded \emph{uplink} symbol-error rate (SER) for a SIMO uplink with $B=16$ BS antennas and transmission over $K=16$ time slots. \algname{} and \apalgname{} achieve near-optimal SER performance (close to that of MRC detection with perfect CSIR) and exhibit a performance similar to ML-JED. At a SER of $0.1$\%, TASER~\cite{castaneda2016} entails a 1\,dB SNR loss for  QPSK, while  MRC with conventional CHEST suffers more than 3\,dB SNR loss for BPSK and QPSK. \rev{For \algname{} and \apalgname{}, the curves show the floating-point performance whereas the markers show the fixed-point performance of our golden models.}}
\label{fig:16x17ul}
\end{figure*}
\begin{figure*}[tp]
\centering
\subfigure[BPSK, Rayleigh fading]{\includegraphics[width=0.325\textwidth]{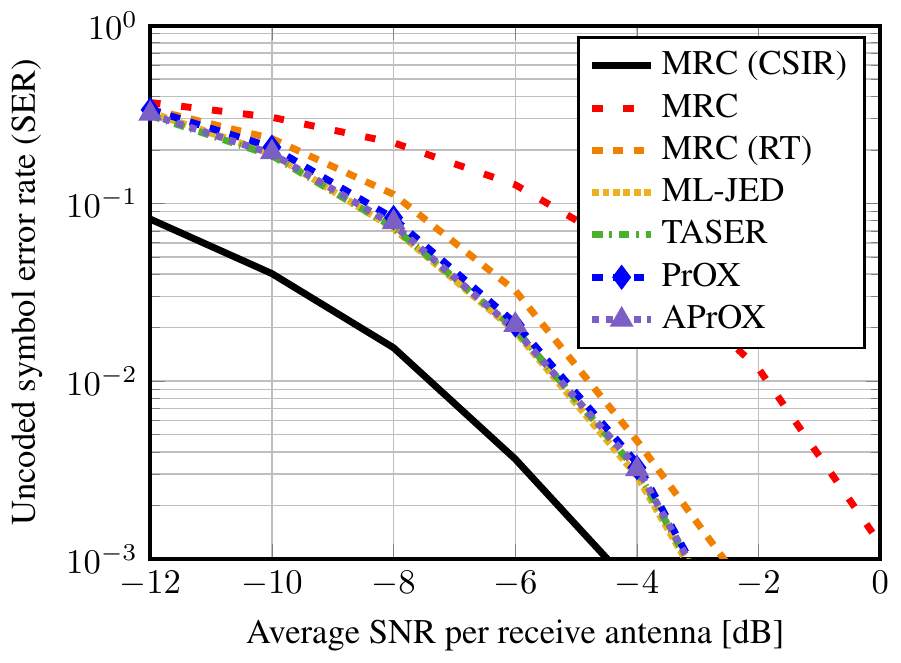}\label{fig:16x17bpskdl_er}}
\subfigure[QPSK, Rayleigh fading]{\includegraphics[width=0.325\textwidth]{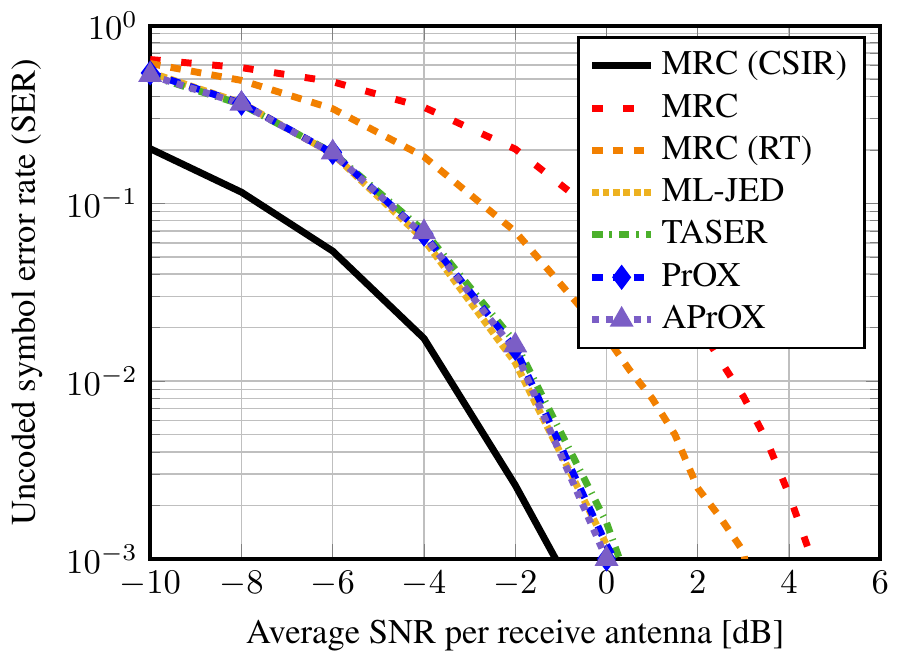}\label{fig:16x17qpskdl_er}}
\subfigure[QPSK, LoS]{\includegraphics[width=0.325\textwidth]{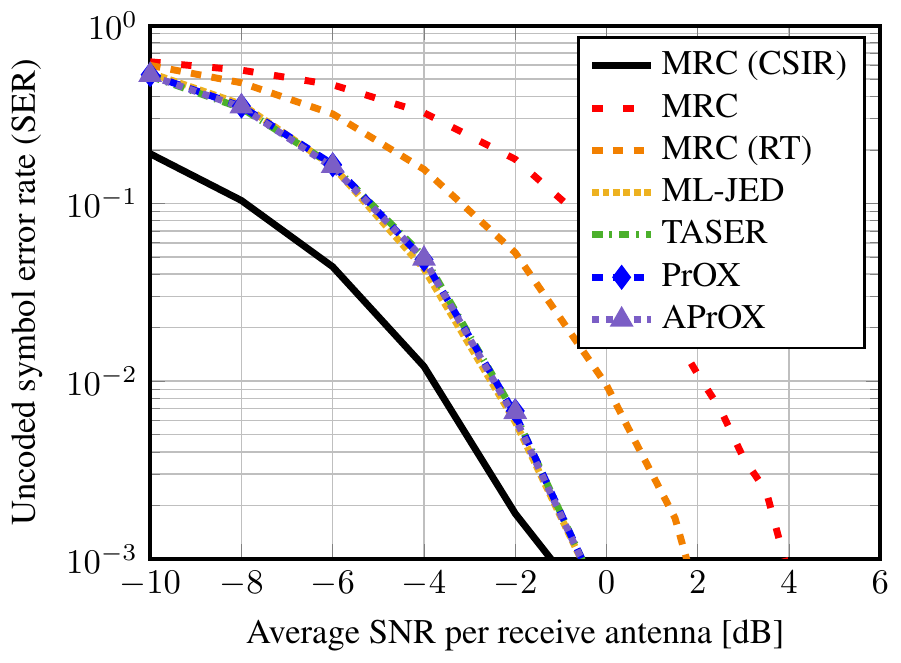}\label{fig:16x17qpsklosdl_er}}
\caption{Uncoded \emph{downlink} symbol-error rate (SER) for a SIMO uplink with $B=16$ and $K=16$. Beamforming with the channel estimate from \algname{}, \apalgname{} and \rev{TASER}~\cite{castaneda2016} achieve near-optimal SER performance (close to that of perfect CSIR) and \rev{a performance} similar to that of ML-JED. At a SER of $0.1$\%, MRC with conventional CHEST suffers a 4\,dB SNR loss for QPSK; \rev{retraining the channel with the estimated uplink data vector can reduce this loss, as in MRC (RT). For \algname{} and \apalgname{}, the curves show the floating-point performance whereas the markers show the fixed-point performance of our golden models.}}
\label{fig:16x17dl}
\end{figure*}
\begin{figure*}[tp]
\centering
\subfigure[BPSK, Rayleigh fading]{\includegraphics[width=0.325\textwidth]{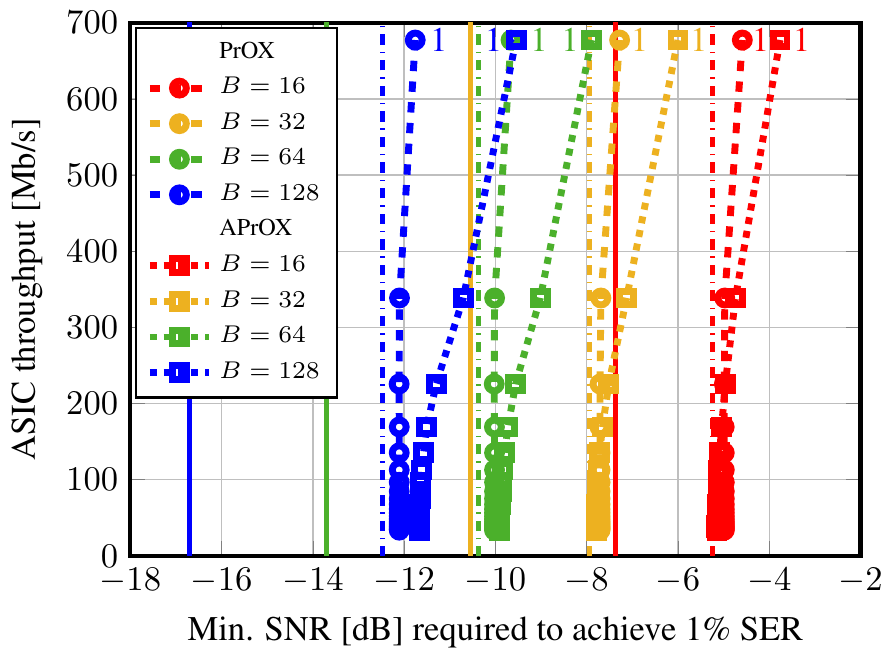}\label{fig:bpsk_tradeoff}}
\subfigure[QPSK, Rayleigh fading]{\includegraphics[width=0.325\textwidth]{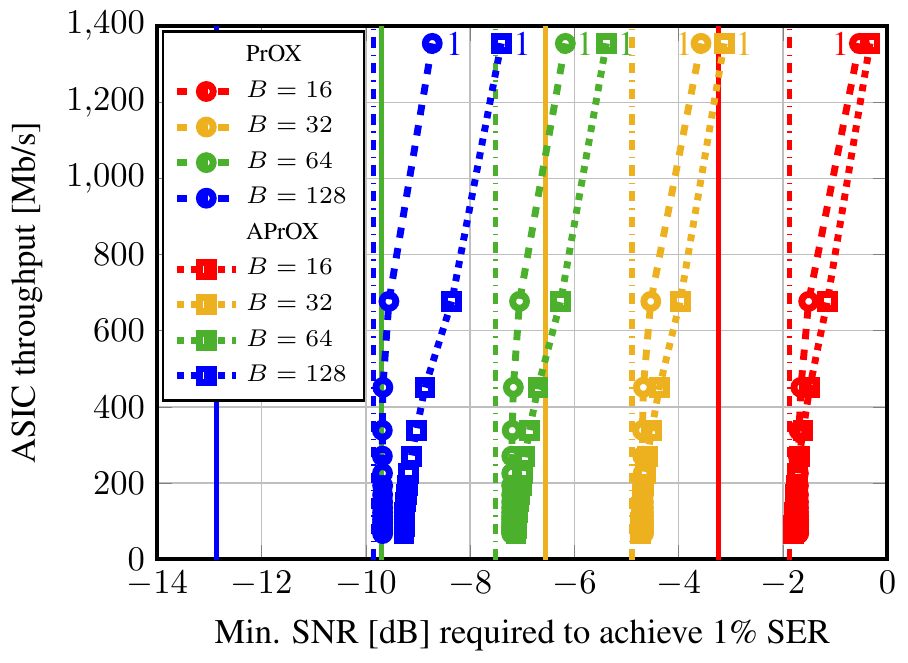}\label{fig:qpsk_tradeoff}}
\subfigure[QPSK, LoS]{\includegraphics[width=0.325\textwidth]{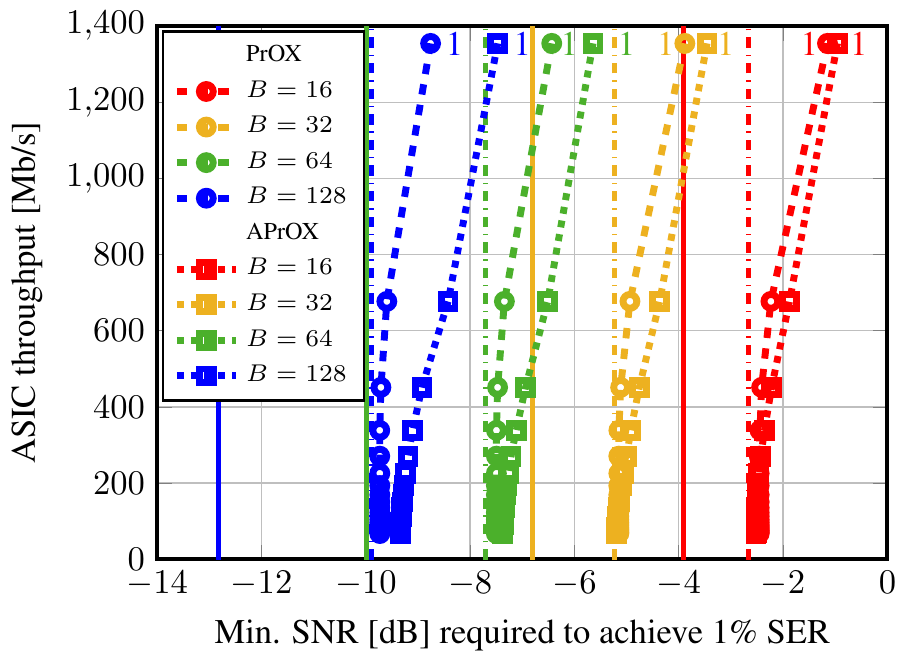}\label{fig:qpsklos_tradeoff}}
\caption{ASIC throughput vs.\ performance trade-off of PrOX and APrOX for $K=16$ time slots. Vertical solid lines represent MRC with CSIR; vertical dashed lines represent ML-JED performance. The numbers next to the markers correspond to the number of iterations~$t_\text{max}$. \rev{The throughput was obtained from post place-and-route ASIC implementation results.} We can see that PrOX requires a smaller number of iterations than APrOX to achieve ML-JED performance.}
\label{fig:tradeoff}
\end{figure*}

\subsection{Error-Rate Performance}
\label{sec:errorrate}

\subsubsection{Uplink Data Detection}
\fref{fig:16x17ul} shows \emph{uplink} symbol error-rate (SER) simulation results for PrOX and APrOX, both running $t_\text{max}=5$ iterations. The simulation results are obtained from $50,000$ Monte-Carlo trials in a $B=16$ BS antenna SIMO system with $K=16$ time slots. We consider both an i.i.d.\ flat Rayleigh block-fading channel model as well as a line-of-sight (LoS) channel with a spherical wave model and a linear BS antenna array with $\lambda/2$-wavelength spacing as described in \cite{tse2005fundamentals}. 
As reference points, we also include  the performance of maximum ratio combining (MRC)-based data detection with both perfect receive-side CSI (denoted by ``CSIR'') and with conventional channel estimation, optimal ML-JED detection using the sphere-decoding algorithm put forward in~\cite{xu2008exact}, and the recently proposed TASER (short for Triangular Approximate SEmidefinite Relaxation) algorithm~\cite{castaneda2016} running for $10$ iterations.
Note that MRC with perfect CSIR is optimal for these scenarios.
However, the assumption of perfect CSIR cannot be realized in practice and one must resort to \rev{channel estimation (CHEST)}, which entails a performance loss of more than 4\,dB at 1\% SER.
PrOX and AProX achieve similar error rate performance. 
Furthermore, both of our algorithms significantly outperform MRC with CHEST and approach near-ML-JED performance but, in stark contrast to ML-JED data detection, at very low computational complexity. 
Our algorithms also outperform TASER, especially for QPSK modulation; see 
\fref{sec:implementationresults} for a hardware comparison. 
We also show the fixed-point performance of the VLSI designs of PrOX. \rev{The markers of PrOX and APrOX correspond to the fixed-point performance of golden models that exactly match the outputs of our VLSI designs, whereas the curves correspond to MATLAB floating-point  performance---evidently, our fixed-point VLSI designs exhibit virtually no implementation loss.}

\subsubsection{Downlink Beamforming}
\fref{fig:16x17dl} shows \emph{downlink} SER simulation results for PrOX and APrOX with the same system and algorithm  parameters. We assume channel reciprocity~\cite{larsson2014massive}, i.e., the downlink channel is the transpose of the uplink channel. Hence, we can use the estimated channel acquired in the uplink for MRC-based beamforming in the downlink. We observe similar trends as for the uplink shown in \fref{fig:16x17ul}. In particular, PrOX, APrOX, and TASER all achieve near-ML-JED performance. MRC and MRC that uses the estimated data vector $\bms^\text{MRC}$ to compute a re-trained channel estimate according to \rev{$\hat\bmh=\bY\hat\bms^\text{MRC}/\|\hat\bms^\text{MRC}\|^2_2$} (referred to as ``RT'') is able to approach our algorithms by 2\,dB and 4\,dB, respectively. 
Hence, we conclude that JED also significantly improves the reliability of data transmission in the downlink. 

\subsubsection{Performance vs.~Complexity Trade-offs}
\fref{fig:tradeoff} shows the trade-offs between the ASIC throughput of PrOX and AProX, and the minimum \rev{signal-to-noise ratio (SNR)} required to achieve $1$\% SER in the uplink for SIMO systems with $K=16$ time slots and a varying number of BS antennas. 
As a reference, we include the performance of MRC with perfect CSIR and that of the optimal ML-JED detector. 
\rev{We observed that by increasing the number of PrOX and APrOX iterations~$t_\text{max}$, the mean squared error (MSE) of the channel estimate is reduced monotonically and the desired SER is achieved at a lower SNR at the expense of reduced throughput.} 
Clearly, PrOX outperforms APrOX for a small number of iterations; this implies that the preprocessing approximation proposed in \fref{sec:preprocessingtrick} entails a small performance loss when a small number of iterations is used.
Note that for BPSK modulation, $t_\text{max}=2$ PrOX~iterations are typically sufficient to reach near-ML-JED performance at an ASIC throughput of $338$\,Mb/s per PrOX instance. For QPSK,  $3$~PrOX iterations are sufficient at an ASIC throughput of $451$\,Mb/s.

\begin{rem}
\rev{For all provided simulation results, we have assumed perfect synchronization and a transmission free of impairments or pilot contamination\footnote{\rev{A straightforward model for pilot contamination would be to add independent Gaussian noise to the received signals---such a model would simply reduce the operating SNR. Without more accurate models, however, it is not clear what the SNR reduction would be in a practical cellular system.}}. Hence, the provided simulation results may not be representative for other system configurations or more realistic communication scenarios. The MATLAB simulator for PrOX used in this paper is available on GitHub: \url{https://github.com/VIP-Group/PrOX}; this enables the interested readers to investigate such aspects in more detail.}
\end{rem}

\subsection{FPGA and ASIC Implementation Results}
\label{sec:implementationresults}

 \begin{table}[t]
  \begin{minipage}[c]{1\columnwidth}
 \vspace{-0.1cm}
    \centering
    \caption{FPGA implementation results of PrOX on a Xilinx Virtex-7 XC7VX690T~for different \algname{} array sizes}
       \label{tbl:implresultstaser}
\renewcommand{\arraystretch}{1.1}       
  \begin{tabular}{@{}lcccc@{}}
  \toprule
  {Array size (${N\!=\!K\!+\!1}$)} & ${N\!=\!5}$ & ${N\!=\!9}$ & ${N\!=\!17}$ & ${N\!=\!33}$ \tabularnewline
  {Time slots  (${K\!=\!N\!-\!1}$) } & ${K\!=\!4}$ & ${K\!=\!8}$ & ${K\!=\!16}$ & ${K\!=\!32}$ \tabularnewline
  \midrule
  {Slices} & 327 & 658 & 1\,491 & 3\,018 \tabularnewline
  {LUTs} & 990 & 1\,991 & 4\,818 & 9\,861 \tabularnewline
  {FFs} & 515 & 989 & 1\,953 & 3\,857 \tabularnewline
  {DSP48s} & 16 & 32 & 64 & 128 \tabularnewline
  Max.\ clock freq. [MHz] & 358 & 341 & 297 & 240 \tabularnewline
  {Min.\ latency [cycles]} & 8 & 12 & 20 & 36 \tabularnewline
  {Max.\ throughput\footnote{Assuming QPSK modulation; for BPSK, the throughput is halved.} [Mb/s]} & 358 &  454 &  475 &  426 \tabularnewline
  {Power estimate\footnote{Statistical power estimation at max.\ clock freq. and 1.0\,V supply voltage.} [W]} & 0.45 & 0.47 & 0.79 & 1.14\tabularnewline
  \bottomrule
  \end{tabular}
  \end{minipage}
  \end{table}
To demonstrate the efficacy of \algname{}, we now provide FPGA implementation results on a Xilinx Virtex-7 XC7VX690T and ASIC implementation results in a 40\,nm CMOS technology. We implemented for different array sizes $N\in\{5,9,17,33\}$ in Verilog on register transfer level (RTL) and hence, our FPGA and ASIC designs support near-ML-JED for $K\in\{4,8,16,32\}$ time slots, respectively. We also provide a comparison to the only other existing FPGA and ASIC designs for JED proposed in the literature~\cite{castaneda2016}. 

\subsubsection{FPGA Implementation Results}
\rev{Our FPGA implementation results were obtained using the Xilinx Vivado Design Suite, and are summarized in \fref{tbl:implresultstaser}.}
As expected, the resource utilization scales nearly linearly with the array size $N$. 
For the $N\in\{5,9,17\}$ arrays, the critical path of PrOX is in the PEs' projection unit; for the largest $N=33$ array, the critical path is in the distribution of the control signals.
  
\fref{tbl:implcomp} compares \algname{} with TASER~\cite{castaneda2016}, which have been implemented on the same FPGA for a SIMO system with $B=128$ BS antennas and communication through $K=8$ time slots.
\algname{} requires significantly fewer resources and lower power than TASER, while achieving a substantially higher throughput.
This makes our design superior to TASER in terms of both hardware-efficiency (measured in throughput per FPGA LUTs) and energy per bit. Concretely, \algname{} is $20\times$ more hardware-efficient and $8\times$ more energy-efficient than TASER for the considered scenario.

 \begin{table}
  \begin{minipage}[c]{1\columnwidth}
            \centering
    \caption{Comparison of JED FPGA designs for a QPSK, $B=128$, $K=8$ large-SIMO system on a Xilinx Virtex-7 XC7VX690T }
       \label{tbl:implcomp}
    \vspace{-1mm}
    \renewcommand{\arraystretch}{1.1}
  \begin{tabular}{@{}lcc@{}}
  \toprule
  {Detection algorithm} & \algname{} &TASER \cite{castaneda2016} \tabularnewline
 {Error-rate performance} & Near ML-JED & Near ML-JED \tabularnewline 
  {Iterations} $t_\text{max}$ & 3 & 3\tabularnewline
  \midrule
  {Slices} & 658 (0.61\,\%) & 4\,350 (4.02\,\%) \tabularnewline
  {LUTs} & 1\,991 (0.46\,\%) & 13\,779 (3.18\,\%) \tabularnewline
  {FFs} & 989 (0.11\,\%) & 6\,857 (0.79\,\%) \tabularnewline
  {DSP48s} & 32 (0.89\,\%) & 168 (4.67\,\%) \tabularnewline
  {Clock frequency [MHz]} & 341 & 225 \tabularnewline
  {Latency [clock cycles]} & 12 & 72  \tabularnewline
  {Throughput [Mb/s]} & 151 & 50 \tabularnewline
  {Power estimate\footnote{Statistical power estimation at max.\ clock freq. and 1.0\,V supply voltage.} [W]} & 0.47 & 1.30 \tabularnewline
  \midrule
  {Throughput/LUTs} & {75\,841} & {3\,629} \tabularnewline
  {Energy/bit [nJ/b]} & {3.09} & {26.0} \tabularnewline
  \bottomrule
  \end{tabular}
  \end{minipage}
  \end{table}

 \setlength{\textfloatsep}{10pt}% Remove \textfloatsep
\begin{figure}[tp]
\centering
\begin{tabular}{cc}
       	\begin{tabular}{@{}c@{}}
	\begin{adjustbox}{valign=t}
		\subfigure[\rule{0pt}{2ex}$N=5$]{\includegraphics[width=0.14 \columnwidth]{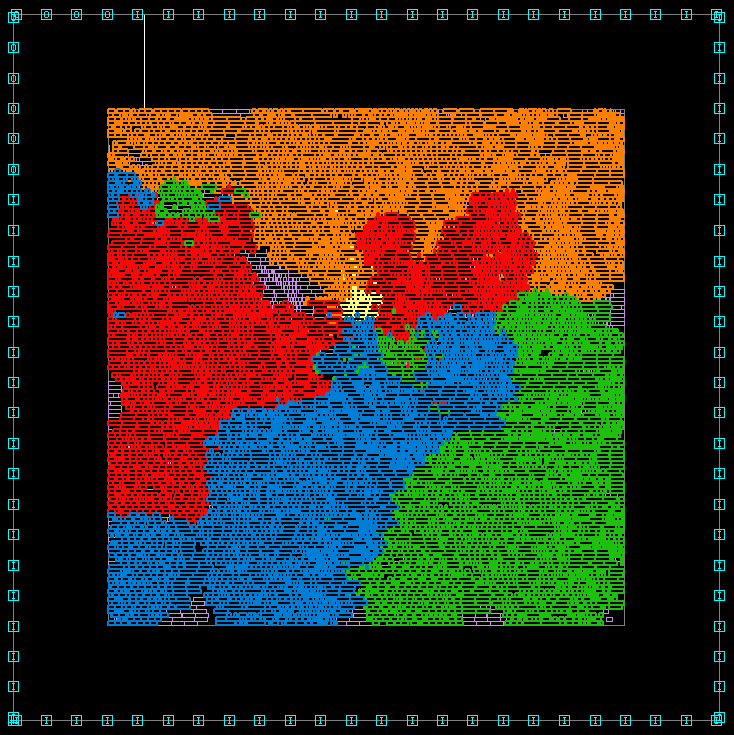}\label{fig:asic_N5}}
	\end{adjustbox}
	\begin{adjustbox}{valign=t}

	\subfigure[\rule{0pt}{2ex}$N=9$]{\includegraphics[width=0.192 \columnwidth]{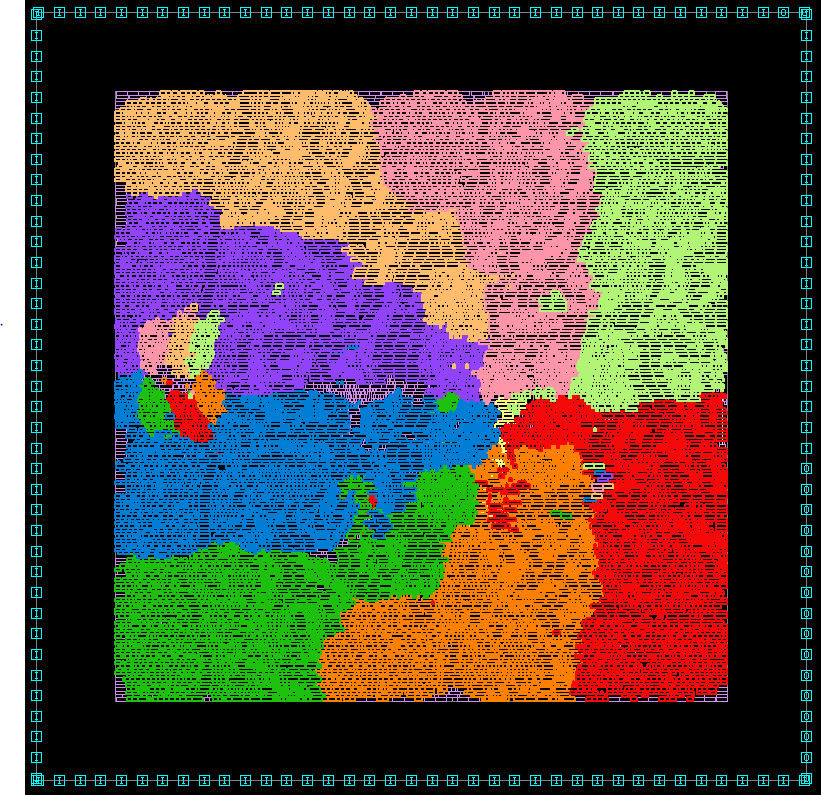}\label{fig:asic_N9}}
	\end{adjustbox}
	~\\
	\begin{adjustbox}{valign=t}
     	\subfigure[\rule{0pt}{2ex}$N=17$]{\includegraphics[width=0.35\columnwidth]{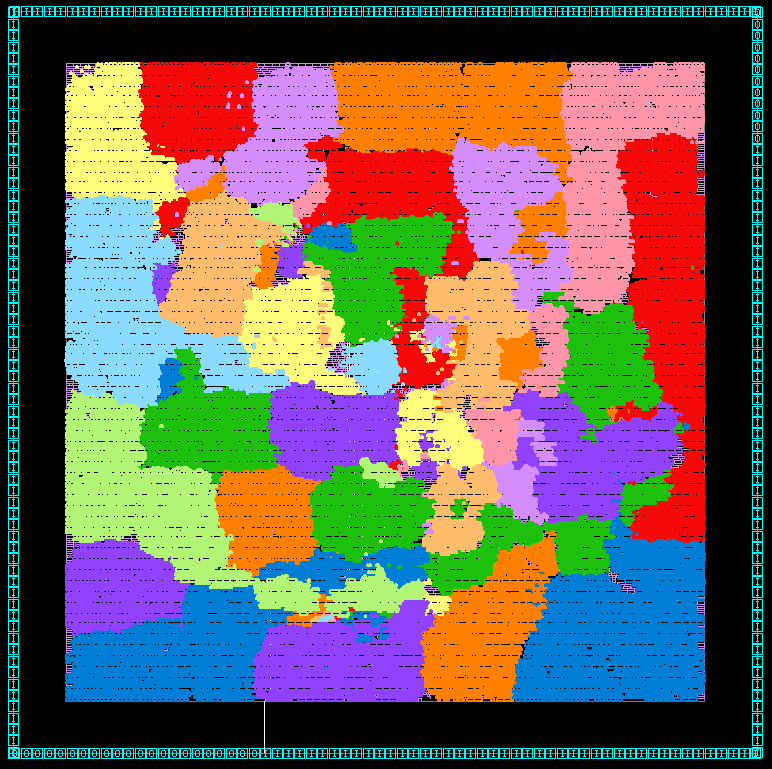}\label{fig:asic_N17}}		
	\end{adjustbox}
	\end{tabular}
	\begin{tabular}{@{}c@{}}
        \begin{adjustbox}{valign=t}
     	\subfigure[\rule{0pt}{2ex}$N=33$]{\includegraphics[width=0.60\columnwidth]{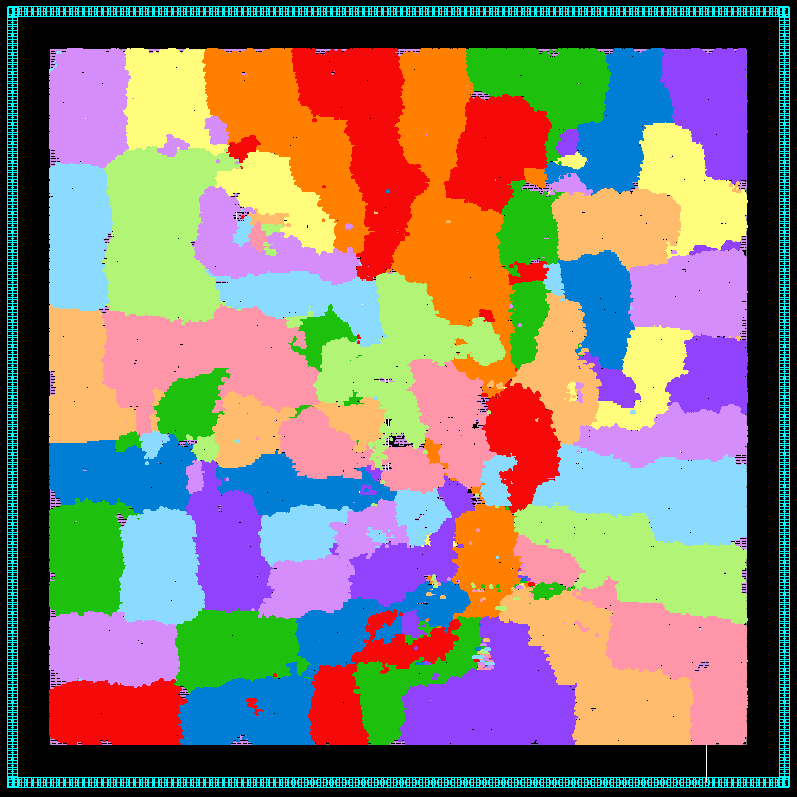}\label{fig:asic_N33}}		
	\end{adjustbox}
        \end{tabular}
\end{tabular}
\caption{Layouts of the PrOX ASIC designs for array sizes $N\in\{5,9,17,33\}$. The PEs of each design were colored differently. One can observe a nearly linear increase in silicon area depending on the array size $N=K+1$.}
\label{fig:barfplot}
\end{figure}

  \begin{table}[tp]
  \begin{minipage}[c]{1\columnwidth}
 \vspace{-0.1cm}
    \centering
    \caption{ASIC implementation results of PrOX in 40\,nm CMOS for different array sizes}
       \label{tbl:asicimplresultstaser}
\renewcommand{\arraystretch}{1.1}       
  \begin{tabular}{@{}lcccc@{}}
  \toprule
  {Array size (${N\!=\!K\!+\!1}$)} & ${N\!=\!5}$ & ${N\!=\!9}$ & ${N\!=\!17}$ & ${N\!=\!33}$ \tabularnewline
  {Time slots  (${K\!=\!N\!-\!1}$) } & ${K\!=\!4}$ & ${K\!=\!8}$ & ${K\!=\!16}$ & ${K\!=\!32}$ \tabularnewline
  \midrule
  Core area [$\mu\text{m}^2$] & 26\,419 & 53\,361 & 132\,903 & 303\,722 \\
  Core density [\%] & 77.24 & 77.39 & 72.96 & 76.38 \\
  Cell area [GE] & 28\,922 & 58\,522 & 137\,428 & 328\,753 \\
  Max.~clock freq. [MHz] & 976 & 942 & 846 & 695 \\
  Min.\ latency [cycles] & 8 & 12 & 20 & 36 \\
  Max.~throughput [Mb/s] & 976 & 1\,256 & 1\,354 & 1\,235  \\
  {Power estimate\footnote{Post place-and-route power estimation at max.~clock freq.~and 1.1\,V.} [mW]} & 10 & 18 & 38 & 58 \\
  \bottomrule
  \end{tabular}
  \end{minipage}
  \end{table}

 \begin{table}[tp]
  \begin{minipage}[c]{1\columnwidth}
            \centering
    \caption{Comparison of JED ASICs for a QPSK, $B=128$, $K=8$ large-SIMO system in 40\,nm CMOS}
       \label{tbl:asicimplcomp}
    \vspace{-1mm}
    \renewcommand{\arraystretch}{1.1}
  \begin{tabular}{@{}lcc@{}}
  \toprule
  {Detection algorithm} & \algname{} &TASER \cite{castaneda2016} \tabularnewline
 {Error-rate performance} & Near ML-JED & Near ML-JED \tabularnewline 
  {Iterations} $t_\text{max}$ & 3 & 3\tabularnewline
  \midrule
  Core area [$\mu\text{m}^2$] & 53\,361 & 482\,677 \\
  Core density [\%] & 77.39 & 68.89\\
  Cell area [GE] & 58\,522 & 471\,238 \\
   Max.~clock freq. [MHz] & 942 & 560 \\
   Latency [clock cycles] & 12 & 24 \\
   Throughput [Mb/s] & 418 & 125 \\
  {Power estimate\footnote{Post place-and-route power estimation at max.~clock freq.~and 1.1\,V.} [mW]} & 18 & 87\tabularnewline
  \midrule
 {Throughput/cell}  & \multirow{2}{*}{7\,153} & \multirow{2}{*}{279}  \\
 {area [b/(s$\times$GE)]} & & \\
 {Energy/bit [pJ/b]} & {43} & {697} \tabularnewline
  \bottomrule
  \end{tabular}
  \end{minipage}
  \end{table}

\subsubsection{ASIC Implementation Results}
\rev{Our ASIC post place-and-route implementation results summarized in \fref{tbl:asicimplresultstaser} were obtained using Synopsys DC and IC compilers with a 40\,nm CMOS standard-cell technology. \fref{fig:barfplot} shows the corresponding ASIC layouts.
For the $N\in\{5,9\}$ arrays, the critical path of PrOX is in the PEs' projection unit; for $N=17$, the critical path is in the multipliers of the MAC unit and, for the largest $N=33$ array, the critical path is in the address generation logic and readout circuitry of the $\hat{\bmg}_k^r$ memories.}

\fref{tbl:asicimplcomp} compares \algname{} with TASER~\cite{castaneda2016}, which have been implemented on the same 40\,nm CMOS technology for a SIMO system with $B=128$  and $K=8$.
\rev{Similar to the FPGA case, the \algname{} ASIC designs require significantly fewer resources and lower power than the TASER ASIC designs, while achieving a substantially higher throughput.}
Concretely, \algname{} is $25\times$ more hardware-efficient (in terms of the throughput per cell area)  and $16\times$ more energy-efficient than TASER for the considered scenario.

We also note that \algname{} exhibits a similar (for BPSK) or better (for QPSK) SER performance than TASER, while using fewer algorithm iterations; see Figures~\ref{fig:16x17ul} and~\ref{fig:16x17dl} for the associated SER results. 
However, while \algname{} is only suitable for JED in large SIMO systems, TASER is also able to perform near-ML data detection in coherent massive multi-user MIMO systems that use conventional pilot-based channel training~\cite{castaneda2016}.
Hence, the error-rate performance and hardware complexity advantages of PrOX over TASER come at the cost of reduced flexibility. 

\rev{
\begin{rem}
While a plethora of VLSI designs exist for data detection in \emph{coherent} small-scale and massive (multi-user) MIMO systems that separate channel estimation from data detection (see \cite{wong2002vlsi, yang2015fifty, burg2005vlsi, studer2010vlsi, jsac07, 5405138, 5560294, 5570931, liao20143, 6725688, Yin2015, Wu2014} and the references therein), TASER  is the only other SIMO-JED VLSI design that has been described in the literature~\cite{castaneda2016}. Furthermore, low-complexity linear methods that are commonly used for conventional (multi-user) MIMO data detection cannot be used in the JED scenario (as briefly discussed in \fref{sec:JEDdetails}). Hence, we limited our comparisons in \fref{tbl:implcomp} and \fref{tbl:asicimplcomp} to \algname{} and TASER.
\end{rem}
}

%%%%%%%%%%%%%%%%%%%%%%%%%%%%%%%%%%%%%%%%%%%%%%%%%%%%

\section{Conclusions}
\label{sec:conclusions}
We have proposed a novel joint channel estimation and data detection (JED) algorithm, referred to as PRojection Onto conveX hull (PrOX). Our algorithm builds upon biconvex relaxation (BCR)~\cite{shah2016biconvex}, which enables near-maximum-likelihood (ML) JED performance at low computational complexity  for large SIMO systems with constant-modulus constellations.
We have provided theoretical convergence guarantees for PrOX and introduced an approximation that significantly reduces the preprocessing complexity.  
To demonstrate the effectiveness of our algorithm, we have proposed a VLSI architecture that consists of a linear array of processing elements (PEs). Our architecture  enables high-throughput near-ML-JED at low silicon area and power consumption.
We have provided FPGA and ASIC implementation results, which demonstrate that PrOX significantly outperforms the only other existing JED design~\cite{castaneda2016} in terms of  hardware- and energy-efficiency as well as error-rate performance.
More generally, \algname{} constitutes a first step towards hardware accelerators that are able to find approximate solutions to problems that resemble the famous MaxCut problem in a hardware-efficient manner. 

There are many avenues for future work. 
An extension of PrOX to compute soft-outputs in the form of log-likelihood ratios (LLRs) is part of ongoing work. Furthermore, 
a theoretical error-rate performance analysis of PrOX is a challenging open research problem. 
\rev{Finally, while some algorithms exist for JED in the MIMO case and for general (non-constant modulus) constellation sets~\cite{xu2008exact,alshamary2016efficient}, they are either computationally complex or exhibit a considerable loss with respect to the optimal ML-JED performance. The development of efficient JED algorithms that can be implemented as high-throughput VLSI designs for large (multi-user) MIMO systems with arbitrary constellations is part of ongoing work.}

%%%%%%%%%%%%%%%%%%%%%%%%%%%%%%%%%%%%%%%%%%%%%%%%%%%%

\appendices

%=========================================================%
\section{Proof of \fref{thm:convergence1}}
\label{app:convergence1}

We need a closed form expression for the gradient so that we can say something about its convergence. We have
 $$\partial_\bmq f(\bmq^{(t)},\bms^{(t)}) = -\bY^T\bY \bmq^{(t)} + \alpha(\bmq^{(t)}-\bms^{(t)}).$$
From the optimality condition for the minimization step  \eqref{minq}, we have 
 \begin{align*}
  0&=-\bY^T\bY \bmq^{(t)} + \alpha(\bmq^{(t)}-\bms^{(t-1)}) \\
   &= -\bY^T\bY \bmq^{(t)} + \alpha(\bmq^{(t)}-\bms^{(t)}) +\alpha(\bms^{(t)}-\bms^{(t-1)})\\
   &=\partial_\bmq f(\bmq^{(t)},\bms^{(t)}) +\alpha(\bms^{(t)}-\bms^{(t-1)}),
   \end{align*}
 and thus 
 \begin{align} 
 \partial_\bmq f(\bmq^{(t)},\bms^{(t)}) =\alpha(\bms^{(t-1)}-\bms^{(t)}). \label{gradform}
\end{align}

Now that we have a simple representation for the gradient, we can bound its norm. Note that $f$ is strongly convex in $\bmq$ with parameter $\alpha - \|\bY^T\bY\|,$ and so
\begin{align} \label{strong1}
f(\bmq,\bms) - f(\bmq_\bms,\bms) \ge \frac{\alpha - \|\bY^T\bY\|}{2}\|\bmq- \bmq_s\|^2_2
\end{align}
  where $\bmq_\bms = \argmin_\bmq f(\bmq,\bms).$  Likewise, $f$ is strongly convex in $\bms$ with parameter $\alpha-\beta,$ and so
\begin{align} \label{strong2}
f(\bmq,\bms) - f(\bmq,\bms_\bmq) \ge \frac{\alpha - \beta}{2}\|\bms- \bms_\bmq\|^2_2
\end{align}
where $\bms_\bmq = \argmin_\bms f(\bmq,\bms).$

If we choose $\bmq=\bmq^{(t-1)}$ and $\bms=\bms^{(t-1)}$ in \eqref{strong1}, then $\bmq_s=\bmq^{(t)}$ and we have
\begin{multline}
f(\bmq^{(t-1)},\bms^{(t-1)}) - f(\bmq^{(t)},\bms^{(t-1)}) \\
\ge \frac{\alpha - \|\bY^T\bY\|}{2}\|\bmq^{(t-1)}- \bmq^{(t)}\|^2_2.
\end{multline}
Similarly, from \eqref{strong2} we get
$$f(\bmq^{(t)},\bms^{(t-1)}) - f(\bmq^{(t)},\bms^{(t)}) \ge \frac{\alpha - \beta}{2}\|\bms^{(t-1)}- \bms^{(t)}\|^2_2.$$
Adding these two inequalities yields
\begin{align*}
f(&\bmq^{(t-1)},\bms^{(t-1)}) - f(\bmq^{(t)},\bms^{(t)}) \\
 &\ge  \frac{\alpha - \|\bY^T\bY\|}{2}\|\bmq^{(t-1)}- \bmq^{(t)}\|^2_2
   +\frac{\alpha - \beta}{2}\|\bms^{(t-1)}- \bms^{(t)}\|^2_2.
 \end{align*}
 Summing from $t=1$ to $t=T,$ and observing that the summation on the left telescopes, we get
 \begin{align} \label{summed}
 f(\bmq^{(0)},\bms^{(0)}) &- f(\bmq^{(T)},\bms^{(T)}) \\
 \ge& \sum_{t=1}^T  \frac{\alpha - \|\bY^T\bY\|}{2}\|\bmq^{(t-1)}- \bmq^{(t)}\|^2_2 \\
 &  +\frac{\alpha - \beta}{2}\|\bms^{(t-1)}- \bms^{(t)}\|^2_2. \nonumber
 \end{align}
Because the iterates remain bounded and $f$ is continuous, the left-hand side of \eqref{summed} is bounded from above. We can conclude that $\|\bms^{(t-1)}- \bms^{(t)}\|\to 0,$  and from \eqref{gradform}, we see that the residuals converge as well.  

Finally, because the sub-gradients of $f$ depend continuously on $\bms$ and $\bmq$, any limit point of the sequence of iterates must have zero sub-gradients, and is thus a stationary point.
 
 %=========================================================%
\section{Proof of \fref{thm:convergence2}}
\label{app:convergence2}

Let $(\bmq\opt{},\bms\opt{} )$ denote a non-zero stationary point of~$f$ that lies in the interior of $\setC_\setO.$  Such a point satisfies the first-order conditions
 \begin{align}
-\bY^T\bY \bmq\opt{}+\alpha (\bmq\opt{}-\bms\opt{}) &= 0 \label{opt_q}\\
\alpha(\bms\opt{} - \bmq\opt{}) - \beta \bms \opt{} &= 0. \label{opt_s}
 \end{align}
From \eqref{opt_s}, we see that
$\bmq\opt{} = \frac{\alpha-\beta}{\alpha} \bms\opt{}.$  Plugging this result into~\eqref{opt_q} yields
$$-\bY^T\bY  \frac{\alpha-\beta}{\alpha} \bms\opt{}+(\alpha-\beta)\bms\opt{}-\alpha\bms\opt{} = 0,$$
which re-arranges to 
$$\bY^T\bY  \bms\opt{}= \frac{-\alpha\beta}{\alpha-\beta}  \bms\opt{}.$$
Since $\bms\opt{}$ is non-zero, this shows that $\bms\opt{}$ is an eigenvector of $\bY^T\bY$ with negative eigenvalue.  This is impossible because the Gram matrix $\bY^T\bY$ is positive semi-definite.  It follows that $(\bmq\opt{},\bms\opt{} )$ cannot satisfy the first-order conditions \eqref{opt_q} and \eqref{opt_s}, and thus, cannot lie in the interior of $\setC_\setO.$ 

%=========================================================%
\section{Proof of \fref{thm:preprocessingapprox}}
\label{app:preprocessingapprox}

Since $\|\bG\|<\alpha$, we have  the Neumann series expansion~\cite{Stewart1998} in~\fref{eq:neumannseries}. Hence, we can bound the approximation error from above as follows:
\begin{align*}
& \|(\bI-\alpha^{-1}\bG)^{-1} - (\bI + \alpha^{-1}\bG) \| = \left\| \sum_{k=2}^\infty (\alpha^{-1}\bG)^k \right\| \\
&\qquad  \overset{\text{(a)}}{\leq} \sum_{k=2}^\infty \left\|  (\alpha^{-1}\bG)^k \right\| 
\overset{\text{(b)}}{\leq} \sum_{k=2}^\infty \left\|  \alpha^{-1}\bG\right\|^k  \\
& \qquad\overset{\text{(c)}}{=} \sum_{k=0}^\infty \left\|  \alpha^{-1}\bG \right\|^k - 1 - \|\alpha^{-1}\bG\| \\
& \qquad \overset{\text{(d)}}{=} \frac{1}{1-\|\alpha^{-1}\bG\|} - 1 - \|\alpha^{-1}\bG\|  
\overset{\text{(e)}}{=}  \frac{\|\alpha^{-1}\bG\|^2}{1-\|\alpha^{-1}\bG\|}.
\end{align*}
Here, step (a) follows from the triangle inequality, step (b) from the fact that the spectral norm is a consistent matrix norm, step (c) is a result of basic arithmetic manipulations, step (d) follows from geometric series expansions, and step (e) is the same expression in simplified form. We note that the proof continues to hold for any consistent matrix norm. 

%%%%%%%%%%%%%%%%%%%%%%%%%%%%%%%%%%%%%%%%%%%%%%%%%%%%

%
\section*{Acknowledgments}
The authors would like to thank S.~Jacobsson and G.~Durisi for discussions on line-of-sight channels. We also thank C.~Jeon and G.~Mirza for discussions on the input-cyclic MVP engine. 
\rev{The work of O.~Casta\~neda and C.~Studer was supported by Xilinx, Inc.\ and by the US National Science Foundation (NSF) under grants ECCS-1408006, CCF-1535897,  CAREER CCF-1652065, and CNS-1717559. The work of T.~Goldstein was supported by the US NSF under grant CCF-1535902 and by the US Office of Naval Research under grant \mbox{N00014-17-1-2078}.}

\balance

%%%%%%%%%%%%%%%%%%%%%%%%%%%%%%%%%%%%%%%%%%%%%%%%%%%%

\balance

% that's all folks

\begin{thebibliography}{10}
\providecommand{\url}[1]{#1}
\csname url@samestyle\endcsname
\providecommand{\newblock}{\relax}
\providecommand{\bibinfo}[2]{#2}
\providecommand{\BIBentrySTDinterwordspacing}{\spaceskip=0pt\relax}
\providecommand{\BIBentryALTinterwordstretchfactor}{4}
\providecommand{\BIBentryALTinterwordspacing}{\spaceskip=\fontdimen2\font plus
\BIBentryALTinterwordstretchfactor\fontdimen3\font minus
  \fontdimen4\font\relax}
\providecommand{\BIBforeignlanguage}[2]{{%
\expandafter\ifx\csname l@#1\endcsname\relax
\typeout{** WARNING: IEEEtran.bst: No hyphenation pattern has been}%
\typeout{** loaded for the language `#1'. Using the pattern for}%
\typeout{** the default language instead.}%
\else
\language=\csname l@#1\endcsname
\fi
#2}}
\providecommand{\BIBdecl}{\relax}
\BIBdecl

\bibitem{CGC17}
O.~{Casta\~neda}, T.~Goldstein, and C.~Studer, ``{FPGA} design of
  low-complexity joint channel estimation and data detection for large {SIMO}
  wireless systems,'' in \emph{Proc. IEEE Int. Symp. Circuits Syst. (ISCAS)},
  May 2017, pp. 128--131.

\bibitem{Marzetta2010}
T.~L. Marzetta, ``Noncooperative cellular wireless with unlimited numbers of
  base station antennas,'' \emph{{IEEE} Trans. Wireless Commun.}, vol.~9,
  no.~11, pp. 3590--3600, Nov. 2010.

\bibitem{Rusek2012}
F.~Rusek, D.~Persson, B.~K. Lau, E.~G. Larsson, T.~L. Marzetta, O.~Edfors, and
  F.~Tufvesson, ``Scaling up {MIMO}: Opportunities and challenges with very
  large arrays,'' \emph{{IEEE} Signal Process. Mag.}, vol.~30, no.~1, pp.
  40--60, Jan. 2013.

\bibitem{hoydis2013massive}
J.~Hoydis, S.~Ten~Brink, and M.~Debbah, ``Massive {MIMO} in the {UL/DL} of
  cellular networks: How many antennas do we need?'' \emph{IEEE J. Sel. Areas
  Commun.}, vol.~31, no.~2, pp. 160--171, Feb. 2013.

\bibitem{larsson2014massive}
E.~Larsson, O.~Edfors, F.~Tufvesson, and T.~Marzetta, ``Massive {MIMO} for next
  generation wireless systems,'' \emph{IEEE Commun. Mag.}, vol.~52, no.~2, pp.
  186--195, Feb. 2014.

\bibitem{andrews2014will}
J.~G. Andrews, S.~Buzzi, W.~Choi, S.~V. Hanly, A.~Lozano, A.~C. Soong, and
  J.~C. Zhang, ``What will {5G} be?'' \emph{IEEE J. Sel. Areas Commun.},
  vol.~32, no.~6, pp. 1065--1082, Jun. 2014.

\bibitem{LUetal2014}
L.~Lu, G.~Li, A.~Swindlehurst, A.~Ashikhmin, and R.~Zhang, ``An overview of
  massive {MIMO}: Benefits and challenges,'' \emph{IEEE J. Sel. Topics Signal
  Process.}, vol.~8, no.~5, pp. 742--758, Oct. 2014.

\bibitem{Wu2014}
M.~Wu, B.~Yin, G.~Wang, C.~Dick, J.~R. Cavallaro, and C.~Studer, ``Large-scale
  {MIMO} detection for {3GPP LTE:} algorithms and {FPGA} implementations,''
  \emph{IEEE J. Sel. Topics Signal Process.}, vol.~8, no.~5, pp. 916--929, Oct.
  2014.

\bibitem{stoica2003space}
P.~Stoica and G.~Ganesan, ``Space--time block codes: Trained, blind, and
  semi-blind detection,'' \emph{Elsevier Dig. Signal Process.}, vol.~13, no.~1,
  pp. 93--105, Jan. 2003.

\bibitem{vikalo2006efficient}
H.~Vikalo, B.~Hassibi, and P.~Stoica, ``Efficient joint maximum-likelihood
  channel estimation and signal detection,'' \emph{{IEEE} Trans. Wireless
  Commun.}, vol.~5, no.~7, pp. 1838--1845, Jul. 2006.

\bibitem{alshamary2015optimal}
H.~A.~J. Alshamary, M.~F. Anjum, T.~Al-Naffouri, A.~Zaib, and W.~Xu, ``Optimal
  non-coherent data detection for massive {SIMO} wireless systems with general
  constellations: A polynomial complexity solution,'' \emph{arXiv preprint:
  1507.02319}, Jul. 2015.

\bibitem{xu2008exact}
W.~Xu, M.~Stojnic, and B.~Hassibi, ``On exact maximum-likelihood detection for
  non-coherent {MIMO} wireless systems: {A} branch-estimate-bound optimization
  framework,'' in \emph{Proc. IEEE Int. Symp. Inf. Theory (ISIT)}, Jul. 2008,
  pp. 2017--2021.

\bibitem{PhamJED}
T.-H. Pham, Y.-C. Liang, and A.~Nallanathan, ``A joint channel estimation and
  data detection receiver for multiuser {MIMO IFDMA} systems,'' \emph{IEEE
  Trans. Commun.}, vol.~57, no.~6, pp. 1857--1865, Jun. 2010.

\bibitem{prasad2015bayes}
R.~Prasad, C.~R. Murthy, and B.~D. Rao, ``Joint channel estimation and data
  detection in {MIMO-OFDM} systems: A sparse {Bayesian} learning approach,''
  \emph{IEEE Trans. Sig. Proc.}, vol.~63, no.~20, pp. 5369--5382, Oct. 2015.

\bibitem{kofidis2016tensor}
E.~Kofidis, C.~Chatzichristos, and A.~L.~F. de~Almeida, ``Joint channel
  estimation / data detection in {MIMO-FBMC/OQAM} systems - a tensor-based
  approach,'' \emph{arXiv preprint: 1609.09661}, Sep. 2016.

\bibitem{wen2016bayes}
C.-K. Wen, C.-J. Wang, S.~Jin, K.-K. Wong, and P.~Ting, ``{Bayes}-optimal joint
  channel-and-data estimation for massive {MIMO} with low-precision {ADCs},''
  \emph{IEEE Trans. Sig. Proc.}, vol.~64, no.~10, pp. 2541--2556, Dec. 2016.

\bibitem{castaneda2016}
O.~Casta\~neda, T.~Goldstein, and C.~Studer, ``Data detection in large
  multi-antenna wireless systems via approximate semidefinite relaxation,''
  \emph{{IEEE} Trans. Circuits Syst. {I}}, no.~99, pp. 2334--2346, Nov. 2016.

\bibitem{shah2016biconvex}
S.~Shah, A.~Kumar, C.~Castillo, D.~Jacobs, C.~Studer, and T.~Goldstein,
  ``Biconvex relaxation for semidefinite programming in computer vision,''
  \emph{European Conf. Comput. Vision (ECCV)}, pp. 717--735, Sep. 2016.

\bibitem{yang2015fifty}
S.~Yang and L.~Hanzo, ``Fifty years of {MIMO} detection: The road to
  large-scale {MIMOs},'' \emph{IEEE Commun. Surveys Tuts.}, vol.~17, no.~4, pp.
  1941--1988, Sep. 2015.

\bibitem{stoica2003joint}
P.~Stoica, H.~Vikalo, and B.~Hassibi, ``Joint maximum-likelihood channel
  estimation and signal detection for {SIMO} channels,'' in \emph{Proc. IEEE
  Int. Conf. Acoust., Speech, Signal Process. (ICASSP)}, vol.~4, May 2003, pp.
  13--16.

\bibitem{alshamary2016efficient}
H.~A.~J. Alshamary and W.~Xu, ``Efficient optimal joint channel estimation and
  data detection for massive {MIMO} systems,'' in \emph{Proc. IEEE Int. Symp. Inf.
  Theory (ISIT)}, Jul. 2016, pp. 875--879.

\bibitem{burg2005vlsi}
A.~Burg, M.~Borgmann, M.~Wenk, M.~Zellweger, W.~Fichtner, and H.~B\"olcskei,
  ``{VLSI} implementation of {MIMO} detection using the sphere decoding
  algorithm,'' \emph{IEEE J. Solid-State Circuits}, vol.~40, no.~7, pp.
  1566--1577, Jul. 2005.

\bibitem{studer2010vlsi}
C.~Studer, M.~Wenk, and A.~Burg, ``{VLSI} implementation of hard-and
  soft-output sphere decoding for wide-band {MIMO} systems,'' in
  \emph{VLSI-SoC: Forward-Looking Trends in IC and Systems Design}.\hskip 1em
  plus 0.5em minus 0.4em\relax Springer, 2010, pp. 128--154.

\bibitem{schenk2013noncoherent}
A.~Schenk and R.~F.~H. Fischer, ``Noncoherent detection in massive {MIMO}
  systems,'' in \emph{Int. ITG Workshop on Smart Antennas (WSA)}, Apr. 2013,
  pp. 1--8.

\bibitem{yammine2016soft}
G.~Yammine and R.~F. Fischer, ``Soft-decision decoding in noncoherent massive
  {MIMO} systems,'' in \emph{Int. ITG Workshop on Smart Antennas (WSA)}, Mar.
  2016, pp. 1--7.

\bibitem{feng2017noncoherent}
J.~Feng, H.~Gao, T.~Wang, T.~Lv, and W.~Guo, ``A noncoherent differential
  transmission scheme for multiuser massive {MIMO} systems,'' in \emph{Proc. IEEE
  Wireless Commun. Networking Conf. (WCNC)}, Mar. 2017, pp. 1--6.

\bibitem{goldsmith2010robust}
B.~S. Thian and A.~Goldsmith, ``Decoding for {MIMO} systems with imperfect
  channel state information,'' in \emph{Proc. IEEE Global Telecommun. Conf.
  (GLOBECOM)}, Dec. 2010, pp. 1--6.

\bibitem{ghods2017optimally}
R.~Ghods, C.~Jeon, G.~Mirza, A.~Maleki, and C.~Studer, ``Optimally-tuned
  nonparametric linear equalization for massive {MU-MIMO} systems,'' in
  \emph{Proc. IEEE Int. Symp. Inf. Theory (ISIT); arXiv preprint: 1705.02985}, Jun.
  2017.

\bibitem{OFDM2004}
R.~Prasad, \emph{{OFDM} for Wireless Communications Systems}.\hskip 1em plus
  0.5em minus 0.4em\relax Norwood, MA, USA: Artech House, Inc., 2004.

\bibitem{3GPP_TS_36.211_v8.6.0}
``{3GPP TS 36.211} {Evolved} {Universal} {Terrestrial} {Radio} {Access}
  ({E-UTRA}) physical channels and modulation (release 8),'' 3rd Generation
  Partnership Project.

\bibitem{DSteinbergThesis}
D.~Steinberg, ``Computation of matrix norms with applications to robust
  optimization,'' Master's thesis, Technion, Israel, 2005.

\bibitem{ajtai1998shortest}
M.~Ajtai, ``The shortest vector problem in {L2} is {NP}-hard for randomized
  reductions,'' in \emph{Proc. ACM Symp. Theory Comput.}\hskip 1em plus 0.5em
  minus 0.4em\relax ACM, 1998, pp. 10--19.

\bibitem{seethaler2011complexity}
D.~Seethaler, J.~Jald{\'e}n, C.~Studer, and H.~B\"olcskei, ``On the complexity
  distribution of sphere decoding,'' \emph{IEEE Trans. Inf. Theory}, vol.~57,
  no.~9, pp. 5754--5768, Sep. 2011.

\bibitem{Wu2016ocd}
M.~Wu, C.~Dick, J.~R. Cavallaro, and C.~Studer, ``High-throughput data
  detection for massive {MU-MIMO-OFDM} using coordinate descent,'' \emph{{IEEE}
  Trans. Circuits Syst. {I}}, no.~99, pp. 2357--2367, Nov. 2016.

\bibitem{nealboyd2013proximal}
N.~Parikh and S.~Boyd, ``Proximal algorithms,'' \emph{Foundations and Trends
  Optimization}, vol.~1, no.~3, pp. 123--231, Jan. 2013.

\bibitem{Stewart1998}
G.~Stewart, \emph{Matrix Algorithms: Basic decompositions}, 1998.

\bibitem{meinerzhagen2010towards}
P.~Meinerzhagen, C.~Roth, and A.~Burg, ``Towards generic low-power
  area-efficient standard cell based memory architectures,'' in \emph{Proc.
  IEEE Int. Midwest Symp. Circuits Syst. (MWSCAS)}, Aug. 2010, pp. 129--132.

\bibitem{tse2005fundamentals}
D.~Tse and P.~Viswanath, \emph{Fundamentals of wireless communication}.\hskip
  1em plus 0.5em minus 0.4em\relax Cambridge University Press, 2005.

\bibitem{wong2002vlsi}
K.~Wong, C.~Tsui, R.~Cheng, and W.~Mow, ``A {VLSI} architecture of a {K-best}
  lattice decoding algorithm for {MIMO} channels,'' in \emph{Proc. IEEE Int.
  Symp. Circuits Syst. (ISCAS)}, vol.~3, May 2002, pp. 273--276.

\bibitem{jsac07}
C.~Studer, A.~Burg, and H.~B\"olcskei, ``Soft-output sphere decoding:
  Algorithms and {VLSI} implementation,'' \emph{IEEE J. Sel. Areas Commun.},
  vol.~26, no.~2, pp. 290--300, Feb. 2008.

\bibitem{5405138}
C.~H. Liao, T.~P. Wang, and T.~D. Chiueh, ``A 74.8 {mW} soft-output detector
  {IC} for 8$\times$8 spatial-multiplexing {MIMO} communications,'' \emph{IEEE
  J. Solid-State Circuits}, vol.~45, no.~2, pp. 411--421, Feb. 2010.

\bibitem{5560294}
C.~H. Yang, T.~H. Yu, and D.~Markovi{\'c}, ``A 5.8 {mW} {3GPP-LTE} compliant
  8$\times$8 {MIMO} sphere decoder chip with soft-outputs,'' in \emph{Symp.
  VLSI Circuits}, Jun. 2010, pp. 209--210.

\bibitem{5570931}
E.~M. Witte, F.~Borlenghi, G.~Ascheid, R.~Leupers, and H.~Meyr, ``A scalable
  {VLSI} architecture for soft-input soft-output single tree-search sphere
  decoding,'' \emph{IEEE Trans. Circuits and Syst. II}, vol.~57, no.~9, pp.
  706--710, Sep. 2010.

\bibitem{liao20143}
C.-F. Liao, J.-Y. Wang, and Y.-H. Huang, ``A 3.1 {Gb/s} 8$\times$8 sorting
  reduced {K-Best} detector with lattice reduction and {QR} decomposition,''
  \emph{IEEE Trans. VLSI Syst.}, vol.~22, no.~12, pp. 2675--2688, Feb. 2014.

\bibitem{6725688}
C.~Senning, L.~Bruderer, J.~Hunziker, and A.~Burg, ``A lattice reduction-aided
  {MIMO} channel equalizer in 90 nm {CMOS} achieving 720 {Mb/s},'' \emph{IEEE
  Trans. Circuits Syst. I}, vol.~61, no.~6, pp. 1860--1871, Jun. 2014.

\bibitem{Yin2015}
B.~Yin, M.~Wu, J.~Cavallaro, and C.~Studer, ``{VLSI Design of Large-Scale
  Soft-Output MIMO Detection Using Conjugate Gradients},'' in \emph{Proc. IEEE
  Int. Conf. Circuits Syst. (ISCAS)}, May 2015, pp. 1498--1501.

\end{thebibliography}
\end{document}